\documentclass[twocolumn,aps,pra,superscriptaddress,tightenlines]{revtex4-1}
\usepackage{amsmath, amsfonts,amssymb}
\usepackage{comment}
\usepackage{soul}
\usepackage{mathtools}			 % Math capabilities
\usepackage{physics}
\usepackage{graphicx}
\usepackage{epsfig}
\usepackage{comment}
\usepackage{color}
\usepackage{textcomp}
\usepackage[colorlinks,citecolor=blue]{hyperref}
\usepackage{CJKutf8}

\usepackage{natbib}
\renewcommand{\Re}{\mathop{\rm Re}}		% for the real part
\renewcommand{\Im}{\mathop{\rm Im}}		% for the imaginary part
\makeatletter
\newcommand*\bigcdot{\mathpalette\bigcdot@{.5}}
\newcommand*\bigcdot@[2]{\mathbin{\vcenter{\hbox{\scalebox{#2}{$\m@th#1\bullet$}}}}}
\makeatother

\begin{document}
\title{Genuine Multipartite Entanglement induced by a Thermal Acoustic Reservoir}
\author{Qing-Yang Qiu}
\affiliation{School of Physics and Institute for Quantum Science and Engineering, Huazhong University of Science and Technology, Wuhan, 430074, China}
\affiliation{Wuhan institute of quantum technology, Wuhan, 430074, China}

\author{Zhi-Guang Lu}
\affiliation{School of Physics and Institute for Quantum Science and Engineering, Huazhong University of Science and Technology, Wuhan, 430074, China}
\affiliation{Wuhan institute of quantum technology, Wuhan, 430074, China}

\author{ Qiongyi He}
\affiliation{School of Physics, Peking University, Beijing, 100871,  China}

\author{Ying Wu}
\affiliation{School of Physics and Institute for Quantum Science and Engineering, Huazhong University of Science and Technology, Wuhan, 430074,  China}
\affiliation{Wuhan institute of quantum technology, Wuhan, 430074, China}

\author{Xin-You L\"{u}}\email{xinyoulu@hust.edu.cn}
\affiliation{School of Physics and Institute for Quantum Science and Engineering, Huazhong University of Science and Technology, Wuhan, 430074,  China}
\affiliation{Wuhan institute of quantum technology, Wuhan, 430074, China}

\date{\today}% It is always \today, today,
\begin{abstract}
Genuine multipartite entanglement (GME) is not only fundamentally interesting for the study of the quantum-to-classical transition, but also is essential for realizing universal quantum computing and quantum networks. Here we investigate the multipartite entanglement (ME) dynamics in a linear chain of $N$ $LC$ resonators interacting optomechanically with a common thermal acoustic reservoir. By presenting the exact analytical solutions of the system evolution, we predict the periodic generation of non-Gaussian ME, including discrete and continuous variables entanglement. Interestingly, GME is obtained even though the system is in a heat bath. The mechanism relies on a special acoustic environment featuring a frequency comb structure. More importantly, our proposed model also allows for the periodic generation of entangled multipartite cat states, i.e., a typical Greenberger-Horne-Zeilinger state, with high fidelity. This work fundamentally broadens the fields of ME, and has wide applications in implementing thermal-noise-resistant quantum information processing and many-body quantum simulation.
\end{abstract}
%corresponding excitations

\maketitle
Multipartite entanglement (ME), with remarkable complexity and diversity~\cite{PhysRevLett.97.140504,PhysRevA.77.022311,GUHNE20091,Rebic:10,PhysRevLett.108.110501,PhysRevA.98.023823,NatPhotonics.15.7476}, outlines fundamental discrepancies to classical physics and emerges at the interface between quantum information and many-body physics~\cite{Nature409}. It successfully endows fundamental physics such as quantum phase transitions and collective radiation with new interpretations~\cite{PhysRevA.73.010305,PhysRevA.74.042317,PhysRevLett.119.250401,arXiv.12515}. Among various types of ME, genuinely entangled states that are not biseparable with respect to any partition~\cite{PhysRevLett.106.250404,PhysRevA.73.010305,Naturephys9.559562,Naturephys11.167172} exhibit more striking advantages in quantum tasks, including quantum computing using cluster states~\cite{PhysRevA.80.022316,PhysRevLett.103.160401} and multiparty quantum networks~\cite{PhysRevLett.103.020501}.

Normally, the generation of ME is acutely fragile in a practical heat environment due to rapid decoherence~\cite{PhysRevLett.93.230501,PhysRevA.73.032345,PhysRevA.81.064304}. In recent years, counterintuitive phenomena of generating quantum entanglement, with the participation of thermal environments, have been confirmed in theory~\cite{PhysRevLett.96.060407,PhysRevA.73.062306,PhysRevB.77.155420,PhysRevA.89.062307} and also supported by experiments~\cite{PhysRevLett.89.277901,PhysRevLett.109.033602,andp.202100038}. Until now, the deterministic generation of genuine multipartite entanglement (GME) in practical environments remains an outstanding challenge, and then a complete understanding of environment-assisted scalable entanglement generation becomes a significant task, although a counterpart has been studied in bipartite quantum systems~\cite{PhysRevLett.129.203604}.

\begin{figure}
  \centering
  \includegraphics[width=8cm]{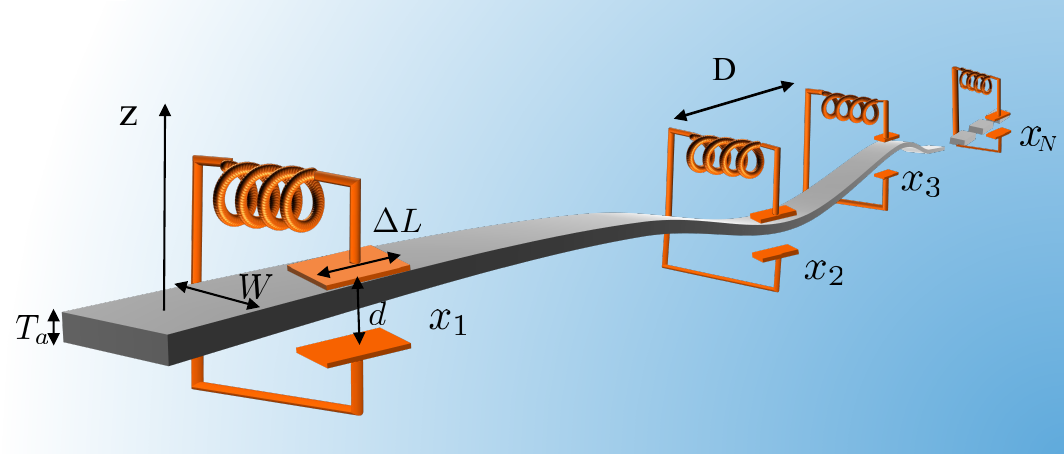}
  \caption{Sketch for a linear chain of $N$ equidistantly distributed $LC$ resonators. The $j$th $LC$ circuit coupled capacitively to an elastic strip, characterized by transverse width $W$ and thickness $T_{a}$ satisfying $T_{a}\ll W \lll L$, at the connecting point $x_{j}$ via metallized segments with lengths $\Delta L$, and the distance between two adjacent coupling points is $D$.  }\label{fig1}
\end{figure}

Here, we explore the non-Gaussian ME dynamics of $N$ $LC$ oscillators that are optomechanically coupled to a one-dimensional (1D) heat elastic strip. By tracing out the phononic degrees of freedom, we analytically determine the time evolution of the system, and then present the periodic generation of discrete and continuous variables non-Gaussian ME (even GME) by utilizing quantum Fisher information (QFI) ~\cite{PhysRevA.88.014301,PhysRevA.99.012304}. The generated GME has strong  robustness against environment temperature near the critical instants, and then it may persist up to a long time under actual thermal environments. Note that, the non-Gaussian entangled states ~\cite{science1250147, nphys3410, PhysRevLett.119.183601} are indispensable resources for universal quantum computing~\cite{PhysRevLett.97.110501}, beyond the Gaussian entanglement states.

Physically, the considered phononic field with a frequency comb structure enables the competing behavior between multipartite entanglement and the dephasing dynamics, leading to a periodic emergence of coherence rephasing. Consequently, the GME, satisfying the causality~\cite{PhysRevLett.89.277901,PhysRevD.97.125002,PhysRevD.100.025004}, could be generated in the array of $N$ $LC$ oscillators under the acoustic heat environments. This property also offers another route for the deterministic generation of various nonclassical states and the long-time storage of many-body quantum resources. As an example, we demonstrate the efficient preparation of an entangled cat state featuring a significant revival with periodicity, when the $N$ resonators are bathed in a common heat bath. The generated ME features essential difference in terms of various aspects (see  Supplemental Material~\cite{SM} and  Refs.~\cite{PhysRevLett.104.020504,PhysRevA.108.022220,Commun328,PhysRevA.64.052311,Commun261,PhysRevA.63.012307,PhysRevA.62.062314,PhysRevLett.87.040401,PhysRevA.67.012108,PhysRevA.65.052112,aravind1997quantum,PhysRevA.100.062329,Aulbach_2010,PhysRevA.81.062347,Kauffman_2002,PhysRevA.98.062335,PhysRevD.107.126005,PhysRevA.85.032314,RevModPhys.81.865,PhysRevLett.78.5022,PhysRevA.65.032314,PhysRevA.61.052306,PhysRevLett.127.040403,JIN2023106155,PhysRevLett.132.151602,PhysRevLett.86.5188,PhysRevLett.99.120503,PhysRevLett.101.130501,PhysRevLett.96.060502,science1250147,PhysRevLett.89.137903,science0070,PhysRevLett.109.230503} therein), when compared with that from the bipartite case.  Our work unveils the periodic many body quantum behavior in a heat phononic bus, which has potential applications in optimizing quantum computing ~\cite{PhysRevA.75.032317,PhysRevLett.127.140501}, quantum information transmission~\cite{PhysRevLett.125.260506,science.adg9210}, and designing more efficient quantum thermal engines~\cite{PhysRevX.5.041011,PhysRevA.101.012315,PhysRevA.108.012433} under  practical environments.

\begin{figure}
  \centering
  \includegraphics[width=6.9cm]{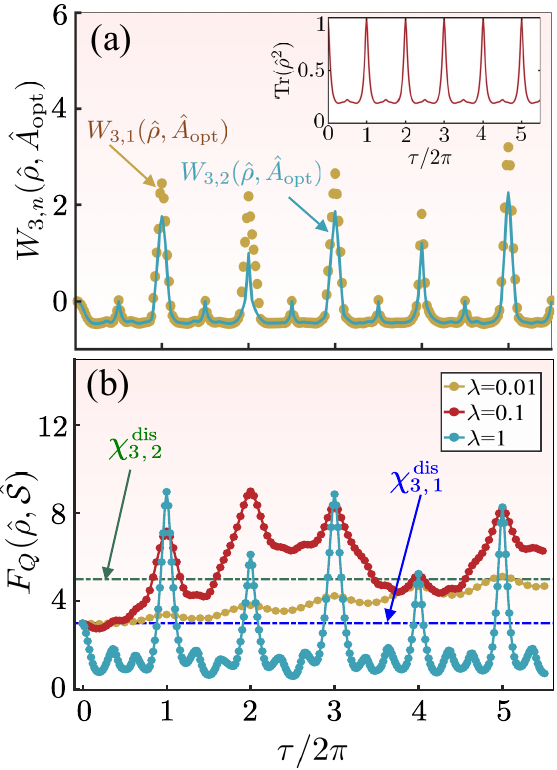}
  \caption{(a) Parameters $W_{n}(\hat{\rho},\hat{A}_{{\rm{opt}}})$ as a function of $\tau/2\pi$ with the initial state $\ket{\alpha}^{\otimes 3}$, where the amplitude $\alpha=0.6$ and $\lambda=1$. The inset depicts the dynamics of $\rm{Tr}(\hat{\rho}^{2})$.  (b) The QFI $F_{Q}(\hat{\rho},\hat{\mathcal{S}})$ as a function of $\tau/2\pi$ for different $\lambda$. The entanglement bounds $\chi^{{\rm dis}}_{3,1}$ (blue dashed line) and $\chi^{{\rm dis}}_{3,2}$ (green dashed line) have been plotted. In (b), the initial state is $\frac{1}{2^{N/2}}(\ket{0}+\ket{1})^{\otimes N}$. Other parameters used here are $\sigma=\pi/4$, $T$ = 0 mK, $\Omega/2\pi$ =15 GHz, $\omega_{1}$ = 50 kHz, $g_{1,1}$ = -10.6 kHz as considered in Refs.\,~\cite{test2, PhysRevLett.129.203604}.
   }\label{fig2}
\end{figure}

\emph{Model and analytical multipartite dynamics.} We consider $N$ identical $LC$ oscillators coupled optomechanically to an elastic mechanical strip with a total length $L$.  As illustrated in Fig.\,\ref{fig1}, the resonators are distributed equidistantly along the elastic strip satisfying $D\gg \Delta L$ and $(N-1)D<L$.  The lower capacitor plate of each $LC$ circuit is assumed to be fixed, and the upper one is attached to the mechanical strip, vibrating up and down along the $z$ axis. There exists a small equilibrium vacuum gap $d\ll W$ between the lower and upper capacitor plates, which gives rise to a balanced capacitance $C=\epsilon_{0} W \Delta L/d$ with $\epsilon_{0}$ the vacuum permittivity. The corresponding flexing mechanical displacement field of the strip, denoted as $u_{z}(t,x)$, is crucial in determining the capacitance of the $k$th $LC$ circuit located at $x_{k}$. The boundary conditions in our case read $u_{z}(t,x=0)=u_{z}(t,x=L)=0$ by assuming a sufficiently large tensile force $F$ is applied at both ends of the strip~\cite{PhysRevA.104.063509}.

By adopting a field theoretic description of the elastic strip, we can obtain the system Hamiltonian
\begin{align}\label{eq1}
\hat{H}=&\sum\limits_{k=1}^{N}\hbar\Omega(\hat{a}^{\dagger}_{k}\hat{a}_{k}+\frac{1}{2})[1+\sum\limits_{j=1}^{\infty}\lambda_{k,j}(\hat{b}_{j}+\hat{b}_{j}^{\dagger})]\nonumber\\
&+\sum\limits_{j=1}^{\infty}\hbar\omega_{j}(\hat{b}^{\dagger}_{j}\hat{b}_{j}+\frac{1}{2}),
\end{align}
where $\hat{a}_{k}$ is the annihilation operator of the $k$th $LC$ oscillator with the resonant frequency $\Omega=1/\sqrt{CL_{a}}$ ($L_{a}$ is the circuit inductances), and $\hat{b}_{j}$ is the annihilation operator for the $j$th mechanical mode of the infinite acoustic reservoir with frequency $\omega_{j}=\frac{j\pi}{L}v_{g}$. Here $v_{g}=\sqrt{F/\rho_{m} W T}$ is the group velocity of phonons propagating along the strip with the corresponding mass density $\rho_{m}$. The optomechanical coupling constants $g_{k,j}\equiv\hbar\Omega\lambda_{k,j}=-\frac{\Omega}{2d}\sqrt{\frac{\hbar}{2m\omega_{j}}}\sin(\frac{\pi j}{L}x_{k})$ ~\cite{PhysRevA.104.063509} are obtained in the limit of a pointlike oscillator, i.e., $\Delta L\rightarrow 0$. Here $m=\rho_{m}WTL/2$ and $x_{k}=[L-D(N-2k+1)]/2$ are the effective mass of the mechanical modes and coupling position of the $k$th $LC$ resonator, respectively. Note that the rotating wave approximation is employed in the derivation of Eq.\,(\ref{eq1}), i.e., the rapidly oscillating terms of the form $\hat{a}_{k}^{2}(\hat{b}_{j}+\hat{b}_{j}^{\dagger})$ and $\hat{a}^{\dagger 2}_{k}(\hat{b}_{j}+\hat{b}_{j}^{\dagger})$ can be safely omitted since the resonant frequency $\Omega$ is much larger than the related coupling $g_{k,j}$.

To obtain the analytical multipartite dynamics, we first expand the initial system state as $\hat{\rho}_{{\rm{in}}}=\sum_{\mathbf{n},\mathbf{n'}}\rho_{\mathbf{n},\mathbf{n'}}(t=0)\ket{\mathbf{n}}\bra{\mathbf{n'}}\otimes\rho_{{\rm{th}}}$. Here the basis $\ket{\mathbf{n}}$ denotes a Fock state that can be marked by a certain set $\mathbf{n}=\{n_{1}n_{2}\cdots n_{N}\}$, and $\rho_{{\rm{th}}}$ describes the thermal state density matrix of the phonon field. The corresponding thermal equilibrium phonon number of $j$th mechanical mode is denoted by $m_{j}=(e^{\hbar\omega_{j}/k_{{\rm{B}}}T}-1)^{-1}$, where $T$ is the environment temperature and $k_{{\rm{B}}}$ is the Boltzmann constant. After tracing out the phononic degrees of freedom, the general analytical expression for the reduced density matrix of $N$ $LC$ oscillators at a time $t$ is given by~\cite{SM}
\begin{widetext}
	\begin{eqnarray}\label{eq2}
	\rho_{\mathbf{n},\mathbf{n'}}(t) \!= &&\exp\bigg\{\!-i\Omega t\sum\limits_{k=1}^{N}(n_{k}-n'_{k})+i\sum\limits_{k=1}^{N}(n_{k}+n'_{k}+1)(n_{k}-n'_{k})\mathcal{P}_{kk}(t)-\sum\limits_{k=1}^{N}(n_{k}-n'_{k})^{2}\mathcal{O}_{kk}(t)
\nonumber\\
&&+\sum\limits_{k>k'}[i(2n_{k}n_{k'}-2n'_{k}n'_{k'}+n_{k}+n_{k'}-n'_{k}-n'_{k'})\mathcal{P}_{kk'}(t)-2(n_{k}-n'_{k})(n_{k'}-n'_{k'})\mathcal{O}_{kk'}(t)]\bigg\}\rho_{\mathbf{n},\mathbf{n'}}(0),
	\end{eqnarray}
\end{widetext}
where
\begin{subequations}
\begin{align}
\mathcal{P}_{kk'}(t)=&\lambda\big\{\tau\Re[\mathrm{Li}_{2}(\xi^{k-k'})-\mathrm{Li}_{2}(-\xi^{N+1-k-k'})]\nonumber\\
&-\frac{1}{2}\Im[\mathrm{Li}_{3}(\delta\xi^{k-k'})-\mathrm{Li}_{3}(-\delta\xi^{N+1-k'-k})\nonumber\\
&\!\!+\!\mathrm{Li}_{3}(\delta\xi^{k'-k})\!-\!\mathrm{Li}_{3}(\!-\delta\xi^{k'+k-N-1})]\big\},\label{eq3a}\\
\mathcal{O}_{kk'}(t)=&\sum_{j=1}^{\infty}\frac{1-\cos(\omega_{j}t)}{\omega_{j}^{2}}g_{k,j}g_{k',j}\coth(\frac{\hbar\omega_{j}}{2k_{B}T}).\label{eq3b}
\end{align}
\end{subequations}
Here we have introduced the $n$-order polylogarithm $\mathrm{Li}_{n}(x)$, and the dimensionless constants $\lambda=\hbar\Omega^{2}/(16md^{2}\omega_{1}^{3})$, $\xi=e^{i\sigma}$, and $\delta=e^{i\tau}$, where $\sigma=\pi D/L$ and $\tau=\omega_{1}t$ are the scaled distance and time, respectively.

The first term in the exponential function appearing on the right-hand side of Eq.\,(\ref{eq2}) is the free evolution of $LC$ resonators. The second and third terms respectively describe the self-Kerr interaction and dephasing of individual oscillators triggered by the acoustic environment. The time-dependent coefficients $\mathcal{P}_{kk'}(t)$ and $\mathcal{O}_{kk'}(t)$ form an effective acoustic communication between $k$th and $k'$th subsystem. Specifically, $\mathcal{P}_{kk'}(t)$ ($k\neq k'$) denotes the cross-Kerr interaction between $k$th and $k'$th oscillators mediated by acoustic phonons that are essentially responsible for the generation of ME. The terms $\mathcal{O}_{kk'}(t)$ suppress the coherence dynamics by forcing the $k$th and $k'$th oscillators to be unentangled. The terms $\mathcal{P}_{kk'}(t)$ also guarantee the type of the extracted entanglement from the environment is time-like and then the causality is satisfied. This is demonstrated theoretically by applying Bernoulli polynomials\,\cite{SM}. It is noteworthy that the competitive interaction between terms $\mathcal{P}_{kk'}$ and $\mathcal{O}_{kk'}$ eventually gives rise to the periodic full rephasing phenomenon, i.e., the exact vanishment of dephasing at time $\tau_{n}=2n\pi$ for $n=1,2,...$, and then the temperature-immune periodic ME generation is available.

Apart from the finite temperature of the acoustic reservoir, the inevitable dissipation also should be included in the realistic quantum systems. Here, we assumed that all of the $LC$ resonators share an identical decay rate $\kappa$ (normally $\kappa\ll\Omega$), and the considered initial system state is the same as $\hat{\rho}_{{\rm{in}}}$. One can approximately solve the corresponding quantum master equation by applying the standard Born iteration procedure~\cite{PhysRevA.55.3042}. After some algebra, the solution of the dissipative dynamics for a linear chain of $N$ $LC$ resonators reads~\cite{SM}
\begin{align}\label{eq4}
\tilde{\rho}_{\mathbf{n},\mathbf{n'}}(t)=&\kappa\sum\limits_{k=1}^{N}\Gamma_{k}(t)\sqrt{(n_{k}+1)(n_{k'}+1)}\rho_{\mathbf{\tilde{n}}_{k},\mathbf{\tilde{n}'}_{k}}(t)
\nonumber\\
&+[1-\frac{\kappa t}{2}\sum\limits_{k=1}^{N}(n_{k}\!+\!n'_{k})]\rho_{\mathbf{n},\mathbf{n'}}(t),
\end{align}
where $\Gamma_{k}(t)=\int_{0}^{t}d\tau e^{2i\sum\limits_{s=1}^{N}[\mathcal{P}_{ks}(\tau)-\mathcal{P}_{ks}(t)](n_{s}-n'_{s})}$ are the $k$-dependent integral coefficients and the notation $\mathbf{\tilde{n}}_{k}=\{n_{1},...,n_{k-1}, n_{k}+1,n_{k+1},..., n_{N}\}$. Since the Eq.\,(\ref{eq4}) merely captures the short time behavior, i.e., $\kappa t\ll 1$, the coupling of the elastic mechanical strip (i.e.,  infinite phonon modes) with its environment can be safely ignored. From now on, the full time evolution of $N$ $LC$ oscillators can be completely determined by Eqs.\,(\ref{eq2}) and (\ref{eq4}), which predicts the generation of ME.

\begin{figure}
  \centering
  \includegraphics[width=8.8cm]{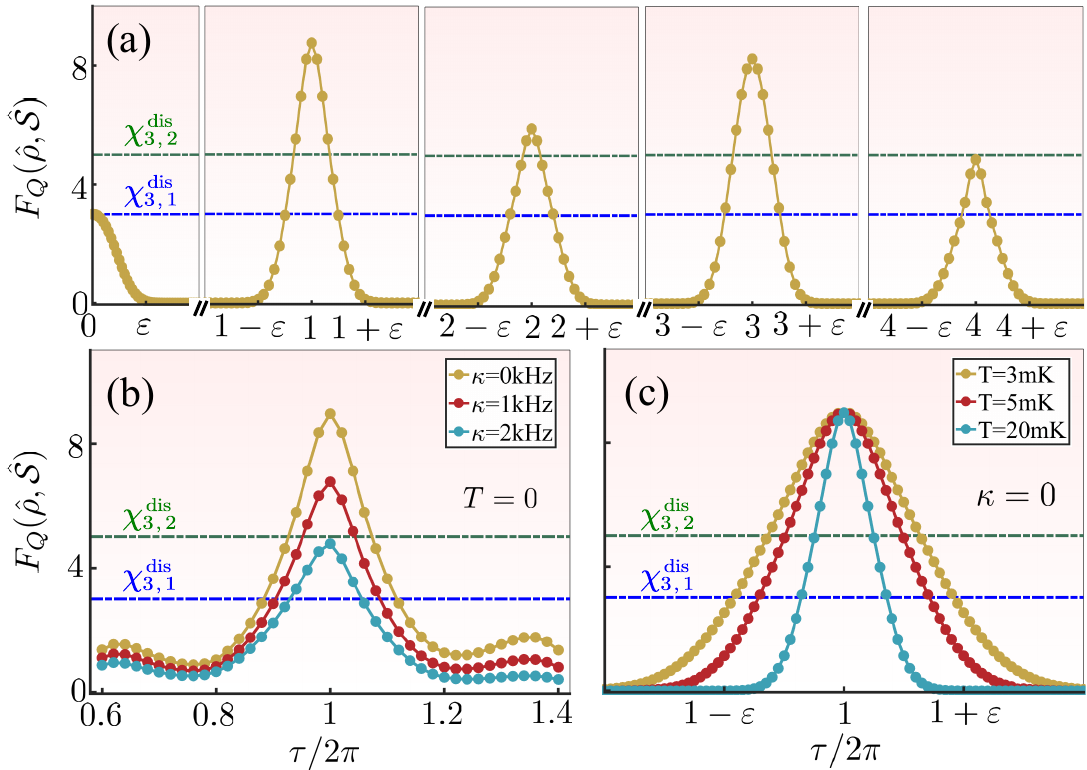}
  \caption{(a) The QFI $F_{Q}(\hat{\rho},\hat{\mathcal{S}})$ vs dimensionless time $\tau/2\pi$ when $T$= 10 mK, $\kappa$ = 0.1 kHz. Evolution of QFI $F_{Q}(\hat{\rho},\hat{\mathcal{S}})$ by (b) increasing the external dissipation $\kappa$ and (c) increasing the environment temperature $T$.  Note that we have fixed respectively $T=0$ and $\kappa=0$ in (b) and (c) for convenience and the time interval $2\pi\varepsilon/ \omega_{1}\approx 62.8$ ns has been introduced. Other parameters used here are the same as in Fig.\,\ref{fig2}.
   }\label{fig3}
\end{figure}

\emph{Periodic ME dynamics.}We assumed first that the $N$ $LC$ oscillators are prepared initially in a separable coherent state $\ket{\alpha}^{\otimes N}$, and the phononic field is in the thermal state $\rho_{{\rm{th}}}$ serving as a thermal environment. For illustration purposes, here we analyze the ME dynamics for $N=3$  based on Eqs.\,(\ref{eq2}) and (\ref{eq4}). The detection of ME for continuous variables~\cite{PhysRevA.94.020101,PhysRevApplied.18.024065, testing1} can be realised by comparing QFI $F_{Q}(\hat{\rho},\hat{A}_{{\rm{opt}}})$ with different bounds $\chi^{{\rm con}}_{N,n}$, where $\hat{A}_{{\rm{opt}}}$ is an optimal system operator making QFI go beyond the bounds as much as possible. Note that the delicate structure of the generated ME can not be detected faithfully by the entanglement measures from bipartite case as illustrated in the Supplemental Material~\cite{SM}. For a given mixed state $\hat{\rho}=\sum\limits p_{k}\ket{k}\bra{k}$ (with $p_{k}>0,\sum_{k}p_{k}=1$), the  QFI $F_{Q}(\hat{\rho},\hat{A} )$ can be evaluated as~\cite{PhysRevLett.72.3439, PhysRevA.85.022322, PhysRevA.88.014301}
\begin{align}\label{eq5}
F_{Q}(\hat{\rho},\hat{A})=2\sum\limits_{p_{k}+p_{l}>0}\frac{(p_{k}-p_{l})^{2}}{(p_{k}+p_{l})}|\bra{k}\hat{A}\ket{l}|^{2}.
\end{align}
Then the efficient criterion is accessible by defining
\begin{align}\label{eq6}
W_{N,n}(\hat{\rho},\hat{A}_{{\rm{opt}}})=F_{Q}(\hat{\rho},\hat{A}_{{\rm{opt}}})-\chi^{{\rm con}}_{N,n}.
\end{align}
For a tripartite system, $W_{3,1}(\hat{\rho},\hat{A}_{{\rm{opt}}})>0$ predicts that there is entanglement between three modes and $W_{3,2}(\hat{\rho},\hat{A}_{{\rm{opt}}})>0$  indicates that the state $\hat{\rho}$ is fully inseparable. The determination of $\hat{A}_{{\rm{opt}}}$ and expressions for different bounds $\chi^{{\rm con}}_{3,n}$ are given in the Supplemental Material ~\cite{SM}.

We simulate the ME dynamics in Figs.\,\ref{fig2}(a) and \ref{fig2}(b) for both continuous and discrete variable system, respectively. In  Fig.\,\ref{fig2}(a), we find that the dynamical behavior of  $W_{3,2}(\hat{\rho},\hat{A}_{{\rm{opt}}})$ is synchronous with that of $W_{3,1}(\hat{\rho},\hat{A}_{{\rm{opt}}})$, which features the periodic positive peaks around $\tau_{n}$.
Therefore, the fully tripartite inseparability between three $LC$ circuits is generated periodically.

Full inseparability of a tripartite state indicates that the state is not biseparable for all of the three possible bipartitions, i.e.,  $i-jk$ for $i \neq j \neq k$. Noticed that the density operator of this state still could be characterized as a mixture
\begin{align}\label{eq7}
\hat{\rho}\!\!=\!\!P_{1}\!\!\sum\limits_{i}\mathcal{F}_{i}^{(1)}\hat{\rho}^{i}_{1,23}\!\!+\!\!P_{2}\!\!\sum\limits_{j}\mathcal{F}_{j}^{(2)}\hat{\rho}^{j}_{2,13}\!\!+\!\!P_{3}\!\!\sum\limits_{k}\mathcal{F}_{k}^{(3)}\hat{\rho}^{k}_{3,12},
\end{align}
where $\sum^{3}_{i=1}P_{i}=1$ , $\sum_{i=1}\mathcal{F}_{i}^{(j)}=1$, and $\hat{\rho}_{i,jk}$ is denoted as $\sum_{\gamma}\eta_{\gamma}\hat{\rho}^{\gamma}_{i}\otimes\hat{\rho}^{\gamma}_{jk}$ for brevity, with $\sum_{\gamma}\eta_{\gamma}=1$.  The entanglement structure of the biseparable form Eq.\,(\ref{eq7}) is conceptually not the GME. However, we find that the evolved states of system are exactly pure states iff  $\tau=\tau_{n}$ [see the inset of Fig.\,\ref{fig2}(a)], due to the elimination of environmental effects. Therefore, the periodic generation of GME is available in our proposal.

We now turn to the case of discrete variable system, where each $LC$ circuit is limited into 2D Hilbert space, i.e., the initial state takes the form of $\frac{1}{2^{N/2}}(\ket{0}+\ket{1})^{\otimes N}$. The ME in this situation can also be quantified through the QFI~\cite{PhysRevA.82.012337,PhysRevA.85.022321,Naturephys3700,Gessner2017entanglement}. Consider a linear observable $\hat{\mathcal{S}}=\frac{1}{2}\sum_{l}\mathbf{e}_{l}\cdot\boldsymbol{\sigma}_{l}$, where $\mathbf{e}_{l}$ is a unit vector on the Bloch sphere, and $\boldsymbol{\sigma}_{l}=(\sigma^{x}_{l},\sigma^{y}_{l},\sigma^{z}_{l})$ is a vector containing the Pauli matrices associated with spin $l$. Then the system hosts at least $(m+1)$-partite entanglement if $F_{Q}(\hat{\rho},\hat{\mathcal{S}})$ fulfills $F_{Q}(\hat{\rho},\hat{\mathcal{S}})>\chi^{{\rm dis}}_{N,n}\equiv sm^{2}+r^{2}$, where $s\equiv\lfloor\frac{N}{m}\rfloor$ and $r=N-sm$. In order to detect ME, we must apply a searching algorithm~\cite{PhysRevA.95.032326} for the real vector $\zeta=(\mathbf{e}_{1},...,\mathbf{e}_{N})^{T}$ to obtain an optimal operator $\hat{\mathcal{S}}$, i.e., maximizing $F_{Q}(\hat{\rho},\hat{\mathcal{S}})$ with $N$ constraints $\mathbf{e}_{k}^{T}\mathbf{e}_{k}=1$. Figure\,\ref{fig2}(b) shows that the system hosts respectively at least two-part and three-part entanglement (i.e., GME), when $F_{Q}(\hat{\rho},\hat{\mathcal{S}})>\chi^{{\rm dis}}_{3,1}$ and $F_{Q}(\hat{\rho},\hat{\mathcal{S}})>\chi^{{\rm dis}}_{3,2}$, respectively.

We are now in a position to analyze the impact of external decay and environment temperature on ME dynamics. In Fig.\,\ref{fig3}(a), we demonstrate that the periodic occurrence of generating GME could persist up to a large times with a finite external decay rate $\kappa$ and environment temperature $T$. Note that, the considered total time evolution in our work approximately reaches to $504\,\mu s$ so that the approximate solution in Eq.\,(\ref{eq4}) is still valid.  Further insights show that increasing $\kappa$ would diminish entanglement, i.e., the peak of QFI is gradually pushed down [Fig.\,\ref{fig3}(b)], and the increase of $T$ narrows the time window during which entanglement is stored among three LC circuits [Fig.\,\ref{fig3}(c)]. Moreover, alternative criteria based on correlation tensors~\cite{PhysRevA.84.062306,PhysRevA.96.052314,Eur.Phys.J.Plus} have been adopted in detecting ME, which agrees well with that utilizing QFI ~\cite{SM}. In contrast to the bipartite case~\cite{PhysRevLett.129.203604}, the effects of decay and environmental temperature on ME carry prominent features in its hierarchical structure and ME-class-determined sensitivity as described detailed in the Supplemental Material~\cite{SM}, by which an adjustable quantum many-body resource is available through dissipation engineering~\cite{Nature1038,Nat660}.
\begin{figure}[t]
  \centering
  \includegraphics[width=8.7cm]{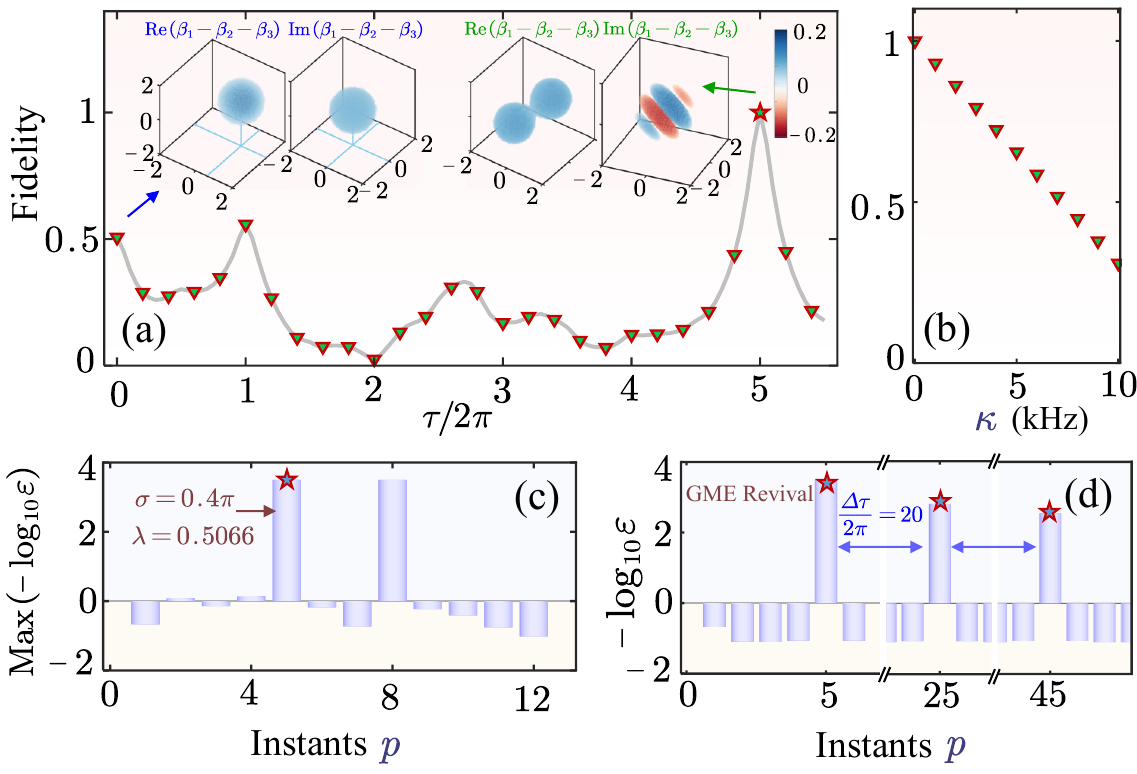}
  \caption{ (a) The fidelity between the states $\ket{\psi_{{\rm cat}}}\bra{\psi_{{\rm cat}}}$ and $\hat{\rho}(t)$ vs scaled time $\tau/2\pi$, with $\sigma=0.4 \pi$ and $\lambda=0.5066$. The insets show the 3D plane-cuts of the measured 6D joint Wigner function for states at $\tau/2\pi=0$ and $\tau/2\pi=5$, respectively. (b) The fidelity between the states $\ketbra{\psi_{{\rm cat}}}$ and $\hat{\rho}(t=t_{p=5})$ as a function of resonators dissipation $\kappa$. (c) The bar graph illustrates the maximal efficiency of generating $\ket{\psi_{{\rm cat}}}$ by scanning the instants $t_{p}$ for $p=0,...,13$. (d) The bar graph of efficiency $-\log_{10}\varepsilon$ as a function of instants $t_{p}$, while fixing the optimal parameters $\sigma=0.4 \pi,\lambda=0.5066$. Other parameters are $N=3, T=0,\alpha=0.7$. }\label{fig4}
\end{figure}

\emph{Generation of entangled multipartite cat states (MCSs).} Of particular interest is the preparation of multi-mode Schrödinger cat states in the linear chain of $N$ $LC$ oscillators. The entangled MCSs, as one of the multipartite quantum resources, are of significance in fundamental tests, such as the generalized Bell-like inequality\,~\cite{PhysRevA.105.052212}, and represents a key prerequisite for practical quantum metrology\,~\cite{science.aay0600}. Inspired by the periodic generation of GME as shown above, here we indicate the potential of preparing cat state with extremely high fidelity. Considering that the resonators are prepared initially in a separable coherent state $\ket{\alpha}^{\otimes N}$, we show that it could deterministically evolve into  $\ket{\psi_{{\rm cat}}}=\frac{1}{\sqrt{2}}(e^{i\pi/4}\ket{\alpha}^{\otimes N} +e^{-i\pi/4} \ket{-\alpha}^{\otimes N})$ at time $t_{p}=2p\pi/\omega_{1}$, when all of the quantities $J_{mn}\equiv \sum\limits_{j=1}^{\infty}\frac{g_{k_{1},j}g_{k_{2},j}}{\omega_{1}\omega_{j}}=\frac{1}{4p}+N_{nm}$ are satisfied for $m,n=1,2,...,N$ with $N_{nm}$ an arbitrary integer (see  Supplemental Material~\cite{SM} and the Refs.~\cite{PhysRevLett.93.266403,PhysRevA.72.041405} therein). Physically, the states $\ket{\psi_{{\rm cat}}}$ can be understood from two heuristic perspectives: a single cat state living in $N$ boxes or $N$ single-cavity cat states that are entangled with each other~\cite{science.aaf2941}. A full quantum-state tomography of the generated MCS can be realized by measuring the joint Wigner function~\cite{sciadv.abn1778}
\begin{align}\label{eq8}
W_{J}(\{\beta\})=(\frac{2}{\pi})^{N}{\rm{Tr}}[\hat{\rho}\prod_{k=1}^{N}\hat{D}_{k}(\beta_{k})\hat{P}_{k}\hat{D}_{k}^{\dagger}(\beta_{k})].
\end{align}
Here $\hat{D}_{k}(\beta_{k})$ refers to the displacement operator of the $k$th cavity mode and $\beta_{k}$ are the complex parameters that construct the coordinates in the joint phase space, and $\hat{P}_{k}=(-1)^{\hat{a}_{k}^{\dagger}\hat{a}_{k}}$ is the corresponding photon number parity operator. Since we focus on the Schr$\rm{\ddot{o}}$dinger's cat that lives in three $LC$ oscillators, the relevant joint Wigner function is limited in 6D phase space $[\Re(\beta_{i}),\Im(\beta_{i})]$ for $i=1,2,3$.

In Fig.\,\ref{fig4}(a), we simulate the dynamics of fidelity $\mathcal{F}[\hat{\rho}_{{\rm cat}},\hat{\rho}(t)]$ between the states $\hat{\rho}_{{\rm cat}}=\ketbra{\psi_{{\rm cat}}}$ and $\hat{\rho}(t)$ for $N=3$, where the fidelity is defined as $\mathcal{F}(\hat{\rho}_{0},\hat{\rho})\equiv{\rm{Tr}}\sqrt{\sqrt{\hat{\rho}}\hat{\rho}_{0}\sqrt{\hat{\rho}}}$. The dimensionless parameter $\lambda$ and scaled distance $\sigma$ are optimized by minimizing the error function $\varepsilon\equiv\sum_{m,n}|e^{2pi\pi J_{mn}}-i|$ as shown in Fig.\,\ref{fig4}(c), from which we find a minimal error of the order $10^{-4}$ at instant $t_{p=5}$.  Such a low conditional error enables us to observe the generation of MCS with a fidelity lager than $99.99\%$ [see pentagram-marked data in Fig.\,\ref{fig4}(a)]. The time window is roughly about $6.91\mu s$, during which $\mathcal{F}[\hat{\rho}_{{\rm cat}},\hat{\rho}(t)]>95\%$. In Fig.\,\ref{fig4}(b), we show the fidelity $\mathcal{F}[\hat{\rho}_{{\rm cat}},\hat{\rho}(t_{p=5})]$ decreases linearly in dissipation $\kappa$.  The 3D plane cuts of the measured 6D joint Wigner function for the generated MCS are plotted in the insets of Fig.\,\ref{fig4}(a), featuring prominent non-classical characteristics by negative valued $W_{J}(\{\beta\})$ in the fringes from space $\Im(\beta_{1}-\beta_{2}-\beta_{3})$. More interestingly, our proposed system also allows the remarkable revival of a three-mode cat state that is verified by the periodic high peaks in the efficiency function $-\log_{10}\varepsilon$. This cycle, for instance, would be $\Delta\tau/2\pi=20$ as shown in Fig.\,\ref{fig4}(d), when fixing optimal parameters obtained in Fig.\,\ref{fig4}(c) at instant $t_{p=5}$.

\emph{Summary and outlook.}---In summary, we have studied analytically the ME dynamics of $N$ $LC$ resonators interacting with a thermal strip in which infinite mechanical oscillators are involved. We found that this system allows the periodic generation of ME including non-Gaussian GME even when the resonators are bathed in a heat bath. The generated multipartite quantum resource, including valuable MCSs, can recover periodically for a long time with potential applications toward thermal-noise-resistant quantum technologies including the quantum internet\,~\cite{nature07127} and quantum key distribution\,~\cite{Nature225}. Following the tremendous progress in experimental physics for superconducting circuits ~\cite{GU20171,Nature810,science3812}, it then would be interesting to investigate the ME in a quantum network where the $LC$ circuits are coupled to a higher dimensional acoustic environment. It is also instructive to design a quantum circuit with a model equivalent to $N$ oscillators immersed in a special acoustic environment featuring frequency comb structure. Also, momentous physics may be explored by combining the present results with platforms pertaining to topological or chiral quantum optics~\cite{Nature1778}.

We thank Prof. Miles P Blencowe and Dr. Qidong Xu for fruitful discussions. This work is supported by the National Key Research and Development Program of China (Grant No. 2021YFA1400700), the National Science Fund for Distinguished Young Scholars of China (Grant No. 12425502),  the National Natural Science Foundation of China (Grant No. 12125402) and the Innovation Program for Quantum Science and Technology (No. 2021ZD0301500). The computation is completed in the HPC Platform of Huazhong University of Science and Technology.
%\bibliography{refermain}
\bibliographystyle{apsrev4-1}

%\let\addcontentsline\oldaddcontentsline% Restore \addcontentsline

%%%%%%%%%% Merge with supplemental materials %%%%%%%%%%
\onecolumngrid

%%%%%%%%%% Prefix a "S" to all equations, figures, tables and reset the counter %%%%%%%%%%
\newcommand\specialsectioning{\setcounter{secnumdepth}{-2}}
\setcounter{equation}{0} \setcounter{figure}{0}

\setcounter{table}{0}
\renewcommand{\theequation}{S\arabic{equation}}
\renewcommand{\thefigure}{S\arabic{figure}}
\renewcommand{\bibnumfmt}[1]{[S#1]}
\renewcommand{\citenumfont}[1]{S#1}
\renewcommand\thesection{S\arabic{section}}
%%%%%%%%%% Prefix a "S" to all equations, figures, tables and reset the counter %%%%%%%%%%
\renewcommand{\baselinestretch}{1.2}

%\renewcommand{\theequation}{S\arabic{equation}}

%%%%%%%%%%%%%%%%%%%%%%%%%%%%%%%%%%%%%%%%%%%%%%%%%%%%%%%%%%%%%%%%%
\newpage

\setcounter{page}{1}\setcounter{secnumdepth}{3} \makeatletter
\begin{center}
{\Large \textbf{ Supplemental Material for\\
        ``Genuine Multipartite Entanglement induced by a Thermal Acoustic Reservoir"}}
\end{center}

\begin{center}
Qing-Yang Qiu$^{1}$, Zhi-Guang Lu$^{1}$,Qiongyi He$^{2}$,Ying Wu$^{1}$, Xin-You L\"{u}$^{1}$
\end{center}

\begin{minipage}[]{16cm}
\small{\it
	\centering $^{1}$School of Physics and Institute for Quantum Science and Engineering, Huazhong University of Science and Technology, and Wuhan Institute of Quantum Technology, Wuhan 430074, China \\
$^{2}$School of Physics, Peking University, Beijing, 100871,  China\\}
\end{minipage}

\vspace{8mm}


\tableofcontents

%%%%%%%%%%%%%%%%%%%%%%%%%%%%%%%%%%%%%%%%%%%%%%
%%%%%%%%%%%%%%%%%%%%%%%%%%%%%%%%%%%%%%%%%%%%%%
\section{ ANALYTICAL TIME EVOLUTION OF \boldmath$N$ LC RESONATORS WITHOUT DISSIPATION}
\setcounter{equation}{0}
\renewcommand\theequation{S\arabic{equation}}
\makeatletter
\renewcommand{\thefigure}{S\@arabic\c@figure}
\makeatother
In this section, we present the analytical derivation, i.e., Eq.(2) in the main text, of the dynamical evolution in the system comprised of $N$ LC circuits that are coupled with a common thermal acoustic reservoir. We remind that the Hamiltonian of our model features standard optomechanical interactions between $N$ cavities and infinite mechanical modes
\begin{align}
\hat{H}=\sum\limits_{k=1}^{N}[\hbar\Omega(\hat{a}^{\dagger}_{k}\hat{a}_{k}+\frac{1}{2})+\sum\limits_{j=1}^{\infty}\hbar g_{k,j}(\hat{a}^{\dagger}_{k}\hat{a}_{k}+\frac{1}{2})(\hat{b}_{j}+\hat{b}_{j}^{\dagger})]
+\sum\limits_{j=1}^{\infty}\hbar\omega_{j}(\hat{b}^{\dagger}_{j}\hat{b}_{j}+\frac{1}{2}),\label{S1}
\end{align}
where the optomechanical coupling constant pertaining to  $k^{\rm{th}}$ LC circuit and $j^{\rm{th}}$ mechanical mode reads\,\cite{PhysRevA.104.063509m}
\begin{align}
g_{k,j}=-\frac{\Omega}{2d}\sqrt{\frac{\hbar}{2m\omega_{j}}}\sin(\frac{\pi j}{L}x_{k}).\label{S2}
\end{align}
Here we have assumed that $N$ oscillators are distributed equidistantly along the elastic strip and the coupling points located at $x_{k}=[L-D(N-2k+1)]/2$.

We first consider a simple initial system-environment state $\ket{\mathbf{n},\{\alpha_{j}\}}$ where the oscillators and acoustic bath are prepared respectively in a product Fork state $\ket{\mathbf{n}}=\ket{\{n_{1}n_{2}\cdots n_{N}\}}=\ket{n_{1}}\otimes\ket{n_{2}}\cdots\otimes\ket{n_{N}}$ and coherent state $\ket{\{\alpha_{j}\}}=\ket{\{\alpha_{1}\alpha_{2}\cdots \alpha_{\infty}\}}=\ket{\alpha_{1}}\otimes\ket{\alpha_{2}}\cdots\otimes\ket{\alpha_{\infty}}$. Here $\ket{n_{k}}$ is Fork state for $k^{\rm{th}}$ LC resonator and $\ket{\alpha_{j}}$ is coherent state for $j^{\rm{th}}$ mechanical mode. The time evolution for such a state governed by Eq.\,(\ref{S1}) is given by
\begin{align}
e^{-\frac{i\hat{H}t}{\hbar}}\ket{\mathbf{n},\{\alpha_{j}\}}=& \hat{U}(t)\ket{\mathbf{n},\{\alpha_{j}\}}\nonumber\\
=&\exp\left(-\frac{it}{\hbar}\left[\sum\limits_{k=1}^{N}\left(\hbar\Omega(\hat{n}_{k}+\frac{1}{2})+\sum\limits_{j=1}^{\infty}\hbar g_{k,j}(\hat{n}_{k}+\frac{1}{2})(\hat{b}_{j}+\hat{b}_{j}^{\dagger})\right)
+\sum\limits_{j=1}^{\infty}\hbar\omega_{j}(\hat{b}^{\dagger}_{j}\hat{b}_{j}+\frac{1}{2})\right]\right)\ket{\mathbf{n},\{\alpha_{j}\}}
\label{S3}.
\end{align}
By applying a unitary transformation with the transformation operator
\begin{align}
\hat{T}=\exp\left(\sum\limits_{k=1}^{N}\sum\limits_{j=1}^{\infty}\frac{g_{k,j}(\hat{n}_{k}+\frac{1}{2})}{\omega_{j}}(\hat{b}_{j}^{\dagger}-\hat{b}_{j})\right),\label{S4}
\end{align}
the mechanical modes obey the transformations:
\begin{align}
\hat{T}\hat{b}_{j}\hat{T}^{\dagger}=&\hat{b}_{j}-\sum_{k=1}^{N}\frac{g_{k,j}(\hat{n}_{k}+\frac{1}{2})}{\omega_{j}},\,\nonumber\\
\hat{T}\hat{b}_{j}^{\dagger}\hat{T}^{\dagger}=&\hat{b}_{j}^{\dagger}-\sum_{k=1}^{N}\frac{g_{k,j}(\hat{n}_{k}+\frac{1}{2})}{\omega_{j}}.\label{S5}
\end{align}
And the time evolution operator $\hat{U}(t)$ is transformed as
\begin{align}
\hat{U'}(t)=&\hat{T}\hat{U}(t)\hat{T}^{\dagger}\,\nonumber\\
=&\exp[-i\Omega t\sum_{k=1}^{N}(\hat{n}_{k}+\frac{1}{2})]\exp(-\sum_{j=1}^{\infty}\frac{i\omega_{j}t}{2})\hat{T}\exp[-it\sum_{k=1}^{N}\sum_{j=1}^{\infty}g_{k,j}(\hat{n}_{k}+\frac{1}{2})(\hat{b}_{j}+\hat{b}_{j}^{\dagger})-it\sum_{j=1}^{\infty}\omega_{j}\hat{b}_{j}^{\dagger}\hat{b}_{j}]\hat{T}^{\dagger}\,\nonumber\\
=&\exp[-i\Omega t\sum_{k=1}^{N}(\hat{n}_{k}+\frac{1}{2})]\exp(-\sum_{j=1}^{\infty}\frac{i\omega_{j}t}{2})\exp(-it\sum_{j=1}^{\infty}\omega_{j}\hat{b}_{j}^{\dagger}\hat{b}_{j})\exp[it\sum_{j=1}^{\infty}\sum_{k,k'=1}^{N}\frac{g_{k,j}g_{k',j}(\hat{n}_{k'}+\frac{1}{2})(\hat{n}_{k}+\frac{1}{2})}{\omega_{j}}].\label{S6}
\end{align}
Then the original time evolution operator $\hat{U}(t)$ is available by applying an inverse transformation to $\hat{U}'(t)$, i.e.,
\begin{align}
\hat{U}(t)=&\hat{T}^{\dagger}\hat{U}'(t)\hat{T}\,\nonumber\\
=&\exp[-i\Omega t\sum_{k=1}^{N}(\hat{n}_{k}+\frac{1}{2})]\exp(-\sum_{j=1}^{\infty}\frac{i\omega_{j}t}{2})\exp[it\sum_{j=1}^{\infty}\sum_{k,k'=1}^{N}\frac{g_{k,j}g_{k',j}(\hat{n}_{k'}+\frac{1}{2})(\hat{n}_{k}+\frac{1}{2})}{\omega_{j}}]\,\nonumber\\
&\times\exp[-\sum_{k=1}^{N}\sum_{j=1}^{\infty}\frac{g_{k,j}(\hat{n}_{k}+\frac{1}{2})}{\omega_{j}}(\hat{b}_{j}^{\dagger}-\hat{b}_{j})]\exp(-it\sum_{j=1}^{\infty}\omega_{j}\hat{b}_{j}^{\dagger}\hat{b}_{j})\exp[\sum_{k=1}^{N}\sum_{j=1}^{\infty}\frac{g_{k,j}(\hat{n}_{k}+\frac{1}{2})}{\omega_{j}}(\hat{b}_{j}^{\dagger}-\hat{b}_{j})]\,\nonumber\\
&\times\exp(it\sum_{j=1}^{\infty}\omega_{j}\hat{b}_{j}^{\dagger}\hat{b}_{j})\exp(-it\sum_{j=1}^{\infty}\omega_{j}\hat{b}_{j}^{\dagger}\hat{b}_{j}).\label{S7}
\end{align}
After some simple algebra, the time evolution operator $\hat{U}(t)$ can be written as a useful form
\begin{align}
\hat{U}(t)=&\exp\left(-i\Omega t\sum\limits_{k=1}^{N}(\hat{n}_{k}+\frac{1}{2})-it\sum\limits_{j=1}^{\infty}\frac{\omega_{j}}{2}+i\sum\limits_{j=1}^{\infty}(\sum\limits_{k=1}^{N}\frac{g_{k,j}}{\omega_{j}}(\hat{n}_{k}+\frac{1}{2}))
(\sum\limits_{k'=1}^{N}\frac{g_{k',j}}{\omega_{j}}(\hat{n}_{k'}+\frac{1}{2}))[\omega_{j}t-\sin(\omega_{j}t)]\right)\nonumber\\
&\times \exp\left(-\sum\limits_{k=1}^{N}\sum\limits_{j=1}^{\infty}\frac{g_{k,j}(\hat{n}_{k}+\frac{1}{2})}{\omega_{j}}[\hat{b}_{j}^{\dagger}(1-e^{-i\omega_{j}t})-\hat{b}_{j}(1-e^{i\omega_{j}t})]\right)\exp(-it\sum\limits_{j=1}^{\infty}\omega_{j}\hat{b}_{j}^{\dagger}\hat{b}_{j}).\label{S8}
\end{align}
Then the Eq.\,(\ref{S3}) can be calculated as
\begin{align}
\hat{U}(t)\ket{\mathbf{n},\{\alpha_{j}\}}=&\exp\left(-i\Omega t\sum\limits_{k=1}^{N}(n_{k}+\frac{1}{2})-it\sum\limits_{j=1}^{\infty}\frac{\omega_{j}}{2}+i\sum\limits_{j=1}^{\infty}(\sum\limits_{k=1}^{N}\frac{g_{k,j}}{\omega_{j}}(n_{k}+\frac{1}{2}))
(\sum\limits_{k'=1}^{N}\frac{g_{k',j}}{\omega_{j}}(n_{k'}+\frac{1}{2}))[\omega_{j}t-\sin(\omega_{j}t)]\right)\nonumber\\
&\times \exp\left(\frac{1}{2}\sum\limits_{k=1}^{N}\sum\limits_{j=1}^{\infty}\frac{g_{k,j}(n_{k}+\frac{1}{2})}{\omega_{j}}(\alpha_{j}^{*}\eta_{j}-\alpha_{j}\eta_{j}^{*})\right)
\ket{\mathbf{n},\{{\alpha_{j}e^{-i\omega_{j}t}-\sum\limits_{k=1}^{N}\frac{g_{k,j}(n_{k}+\frac{1}{2})}{\omega_{j}}\eta_{j}^{*}}\}},\label{S9}
\end{align}
where $\eta_{j}=1-\exp(i\omega_{j}t)$.

Based on the above analytical result Eq.\,(\ref{S9}) , we are able to solve the time evolution of a more general initial system-environment state with the bath is set at a thermal state. Expanding the thermal state density matrix in the coherent basis, this initial state can be described as
\begin{align}
\hat{\rho}_{{\rm{in}}}=\sum\limits_{\mathbf{n},\mathbf{n'}}\rho_{\mathbf{n},\mathbf{n'}}(t=0)\ket{\mathbf{n}}\!\bra{\mathbf{n'}}\otimes\prod_{j}\frac{e^{\beta\hbar\omega_{j}}-1}{\pi}\int d^{2}\alpha_{j}\exp(-|\alpha_{j}|^{2}(e^{\beta\hbar\omega_{j}}-1))\ket{\alpha_{j}}\bra{\alpha_{j}}\label{S10}
\end{align}
with $\beta=1/k_{B}T$ in which $T$ is the environment temperature and $k_{{\rm{B}}}$ is the Boltzmann constant. The system state at time $t$ can be obtained by substituting Eq.\,(\ref{S9}) into Eq.\,(\ref{S10}), i.e.,
\begin{align}
\hat{\rho}(t)=&\sum\limits_{\mathbf{n},\mathbf{n'}}\rho_{\mathbf{n},\mathbf{n'}}(t=0)\ket{\mathbf{n}}\bra{\mathbf{n'}}\nonumber\\
&\times\exp\left(-i\Omega t\sum\limits_{k=1}^{N}(n_{k}-n'_{k})+i\sum\limits_{j=1}^{\infty}\sum\limits_{k,k'=1}^{N}(\frac{g_{k,j}g_{k',j}}{\omega^{2}_{j}}
[(n_{k}+\frac{1}{2})(n_{k'}+\frac{1}{2})-(n'_{k}+\frac{1}{2})(n'_{k'}+\frac{1}{2})][\omega_{j}t-\sin(\omega_{j}t)]\right)\nonumber\\
&\times\prod_{j}\frac{e^{\beta\hbar\omega_{j}}-1}{\pi}\int d^{2}\alpha_{j}\exp(-|\alpha_{j}|^{2}(e^{\beta\hbar\omega_{j}}-1))
\exp\left(\frac{1}{2}\sum\limits_{k=1}^{N}\sum\limits_{j=1}^{\infty}\frac{g_{k,j}(n_{k}-n'_{k})}{\omega_{j}}(\alpha_{j}^{*}\eta_{j}-\alpha_{j}\eta_{j}^{*})\right)\nonumber\\
&\otimes\ket{\{{\alpha_{j}e^{-i\omega_{j}t}-\sum\limits_{k=1}^{N}\frac{g_{k,j}(n_{k}+\frac{1}{2})}{\omega_{j}}\eta_{j}^{*}}\}}\bra{\{{\alpha_{j}e^{-i\omega_{j}t}-\sum\limits_{k=1}^{N}\frac{g_{k,j}(n'_{k}+\frac{1}{2})}{\omega_{j}}\eta_{j}^{*}}\}}.
\label{S11}
\end{align}
After tracing out the phononic degrees of freedom, we obtain the reduced density matrix of the linear chain of $N$ LC resonators
\begin{align}
\hat{\rho}_{\mathbf{n},\mathbf{n'}}(t)=&\sum\limits_{\mathbf{n},\mathbf{n'}}\rho_{\mathbf{n},\mathbf{n'}}(t=0)\ket{\mathbf{n}}\bra{\mathbf{n'}}\nonumber\\
&\times\exp\left(-i\Omega t\sum\limits_{k=1}^{N}(n_{k}-n'_{k})+i\sum\limits_{j=1}^{\infty}\sum\limits_{k,k'=1}^{N}(\frac{g_{k,j}g_{k',j}}{\omega^{2}_{j}}
[(n_{k}+\frac{1}{2})(n_{k'}+\frac{1}{2})-(n'_{k}+\frac{1}{2})(n'_{k'}+\frac{1}{2})][\omega_{j}t-\sin(\omega_{j}t)]\right)\nonumber\\
&\times\prod_{j}\frac{e^{\beta\hbar\omega_{j}}-1}{\pi}\exp\left(-\frac{1}{\omega_{j}^{2}}[(\sum\limits_{k=1}^{N}g_{k,j}(n_{k}+\frac{1}{2}))-(\sum\limits_{k=1}^{N}g_{k,j}(n'_{k}+\frac{1}{2}))]^{2}(1-\cos\omega_{j}t)\right)\nonumber\\
&\int d^{2}\alpha_{j}\exp\left(-|\alpha_{j}|^{2}(e^{\beta\hbar\omega_{j}}-1)
+\sum\limits_{k=1}^{N}\frac{g_{k,j}(n_{k}-n'_{k})}{\omega_{j}}[\alpha_{j}^{*}(1-\exp(i\omega_{j}t))-\alpha_{j}(1-\exp(-i\omega_{j}t))]\right)\nonumber\\
=&\sum\limits_{\mathbf{n},\mathbf{n'}}\rho_{\mathbf{n},\mathbf{n'}}(t=0)\exp\bigg(-i\Omega t\sum\limits_{k=1}^{N}(n_{k}-n'_{k})+i\sum\limits_{j=1}^{\infty}\frac{[\sum\limits_{k}^{N}g_{k,j}(n_{k}+\frac{1}{2})]^{2}-[\sum\limits_{k}^{N}g_{k,j}(n'_{k}+\frac{1}{2})]^{2}}{\omega^{2}_{j}}
[\omega_{j}t-\sin(\omega_{j}t)]\nonumber\\
&-2\sum\limits_{j=1}^{\infty}(\sum\limits_{k=1}^{N}\frac{g_{k,j}}{\omega_{j}}(n_{k}-n'_{k}))^{2}\sin^{2}\frac{\omega_{j}t}{2}\coth(\frac{\beta\hbar\omega_{j}}{2})
\bigg)\ket{\mathbf{n}}\bra{\mathbf{n'}},
\label{S12}
\end{align}
where we have used the identity $\int_{-\infty}^{+\infty}d^{2}\alpha e^{-a|\alpha|^{2}-A\alpha-B\alpha^{*}}=\frac{\pi}{a}e^{AB/a}$ in the above calculation. Note that the summations over infinite mechanical modes emerging in Eq.\,(\ref{S9}) can be given by several simple closed forms
\begin{align}
&t\sum\limits_{j=1}^{\infty}\frac{g_{k,j}^{2}}{\omega_{j}}=\lambda\tau(\frac{\pi^{2}}{6}-\Re[{\rm{Li}}_{2}(-e^{i\sigma(N-2k+1)})]),\label{S13}\\
&t\sum\limits_{j=1}^{\infty}\frac{g_{k,j}g_{k',j}}{\omega_{j}}=\lambda\tau\Re[{\rm{Li}}_{2}(e^{i\sigma(k-k')})-{\rm{Li}}_{2}(-e^{i\sigma(N+1-k-k')})],\label{S14}\\
&\sum\limits_{j=1}^{\infty}\frac{\sin(\omega_{j}t)g_{k,j}^{2}}{\omega^{2}_{j}}=\lambda\Im[{\rm{Li}}_{3}(e^{i\tau})-\frac{1}{2}{\rm{Li}}_{3}(-e^{i[\tau+\sigma(N+1-2k)]})
-\frac{1}{2}{\rm{Li}}_{3}(-e^{i[\tau-\sigma(N+1-2k)]})],\label{S15}\\
&\sum\limits_{j=1}^{\infty}\frac{\sin(\omega_{j}t)g_{k,j}g_{k',j}}{\omega^{2}_{j}}=\frac{\lambda}{2}\Im[{\rm{Li}}_{3}(e^{i[\tau+\sigma(k-k')]})+{\rm{Li}}_{3}(e^{i[\tau-\sigma(k-k')]})
-{\rm{Li}}_{3}(-e^{i[\tau+\sigma(N+1-k-k')]})-{\rm{Li}}_{3}(-e^{i[\tau-\sigma(N+1-k'-k)]})],\label{S16}
\end{align}
where ${\rm{Li}}_{n}(\bigcdot)$ denotes the polylogarithm of order $n$, and we have introduced the notations $\tau=\omega_{1}t,\,\sigma=\pi D/L,\,\lambda=\frac{\hbar\Omega_{b}^{2}}{16md^{2}\omega_{1}^{3}}$ for convenience. Inserting Eqs.\,(\ref{S13})-(\ref{S16}) into Eq.\,(\ref{S12}), then we have
\begin{align}
\rho_{\mathbf{n},\mathbf{n'}}(t)=&\exp\bigg\{-i\Omega t\sum\limits_{k=1}^{N}(n_{k}-n'_{k})+i\sum\limits_{k=1}^{N}[(n_{k}+\frac{1}{2})^{2}-(n'_{k}+\frac{1}{2})^{2}]
\sum\limits_{j=1}^{\infty}\frac{g_{k,j}^{2}t}{\omega_{j}}\nonumber\\
&+2i\sum\limits_{k>k'}^{N}[(n_{k}+\frac{1}{2})(n_{k'}+\frac{1}{2})-(n'_{k}+\frac{1}{2})(n'_{k'}+\frac{1}{2})]\sum\limits_{j=1}^{\infty}\frac{g_{k,j}g_{k',j}}{\omega_{j}}t\nonumber\\
&-i\sum\limits_{k=1}^{N}[(n_{k}+\frac{1}{2})^{2}-(n'_{k}+\frac{1}{2})^{2}]\sum\limits_{j=1}^{\infty}\frac{g_{k,j}^{2}\sin(\omega_{j}t)}{\omega^{2}_{j}}\nonumber\\
&-2i\sum\limits_{k>k'}^{N}[(n_{k}+\frac{1}{2})(n_{k'}+\frac{1}{2})-(n'_{k}+\frac{1}{2})(n'_{k'}+\frac{1}{2})]\sum\limits_{j=1}^{\infty}\frac{g_{k,j}g_{k',j}\sin(\omega_{j}t)}{\omega^{2}_{j}}\nonumber\\
&-\sum\limits_{j=1}^{\infty}\sum\limits_{k,k'=1}^{N}\frac{g_{k,j}g_{k',j}}{\omega^{2}_{j}}(n_{k}-n'_{k})(n_{k'}-n'_{k'})(1-\cos(\omega_{j}t))\coth(\frac{\beta\hbar\omega_{j}}{2})\bigg\}\rho_{\mathbf{n},\mathbf{n'}}(0)\nonumber
\end{align}
\begin{align}
=&\exp\bigg\{\!-i\Omega t\sum\limits_{k=1}^{N}(n_{k}-n'_{k})+i\sum\limits_{k=1}^{N}(n_{k}+n'_{k}+1)(n_{k}-n'_{k})\mathcal{P}_{kk}(t)-\sum\limits_{k=1}^{N}(n_{k}-n'_{k})^{2}\mathcal{O}_{kk}(t)
\nonumber\\
&+\sum\limits_{k>k'}[i(2n_{k}n_{k'}-2n'_{k}n'_{k'}+n_{k}+n_{k'}-n'_{k}-n'_{k'})\mathcal{P}_{kk'}(t)-2(n_{k}-n'_{k})(n_{k'}-n'_{k'})\mathcal{O}_{kk'}(t)]\bigg\}\rho_{\mathbf{n},\mathbf{n'}}(0),\label{S17}
\end{align}
i.e., Eq.(2) in the main text. Note that we have defined the indirect interactions, mediated by acoustic phonons, between $k^{\rm{th}}$ and $k'^{\rm{th}}$ LC resonators as follows:
\begin{align}
\mathcal{P}_{kk'}(t)=&\lambda\big\{\tau\Re[\mathrm{Li}_{2}(e^{i\sigma(k-k')})-\mathrm{Li}_{2}(-e^{i\sigma(N+1-k-k')})]
-\frac{1}{2}\Im[\mathrm{Li}_{3}(e^{i[\tau+\sigma(k-k')]})-\mathrm{Li}_{3}(-e^{i[\tau+\sigma(N+1-k'-k)]})\nonumber\\
&+\mathrm{Li}_{3}(e^{i[\tau-\sigma(k-k')]})-\mathrm{Li}_{3}(-e^{i[\tau-\sigma(N+1-k'-k)]})]\big\},\label{S18}\\
\mathcal{O}_{kk'}(t)=&\sum_{j=1}^{\infty}\frac{1-\cos(\omega_{j}t)}{\omega_{j}^{2}}g_{k,j}g_{k',j}\coth(\frac{\beta\hbar\omega_{j}}{2}).\label{S19}
\end{align}
\begin{figure}
  \centering
  \includegraphics[width=17cm]{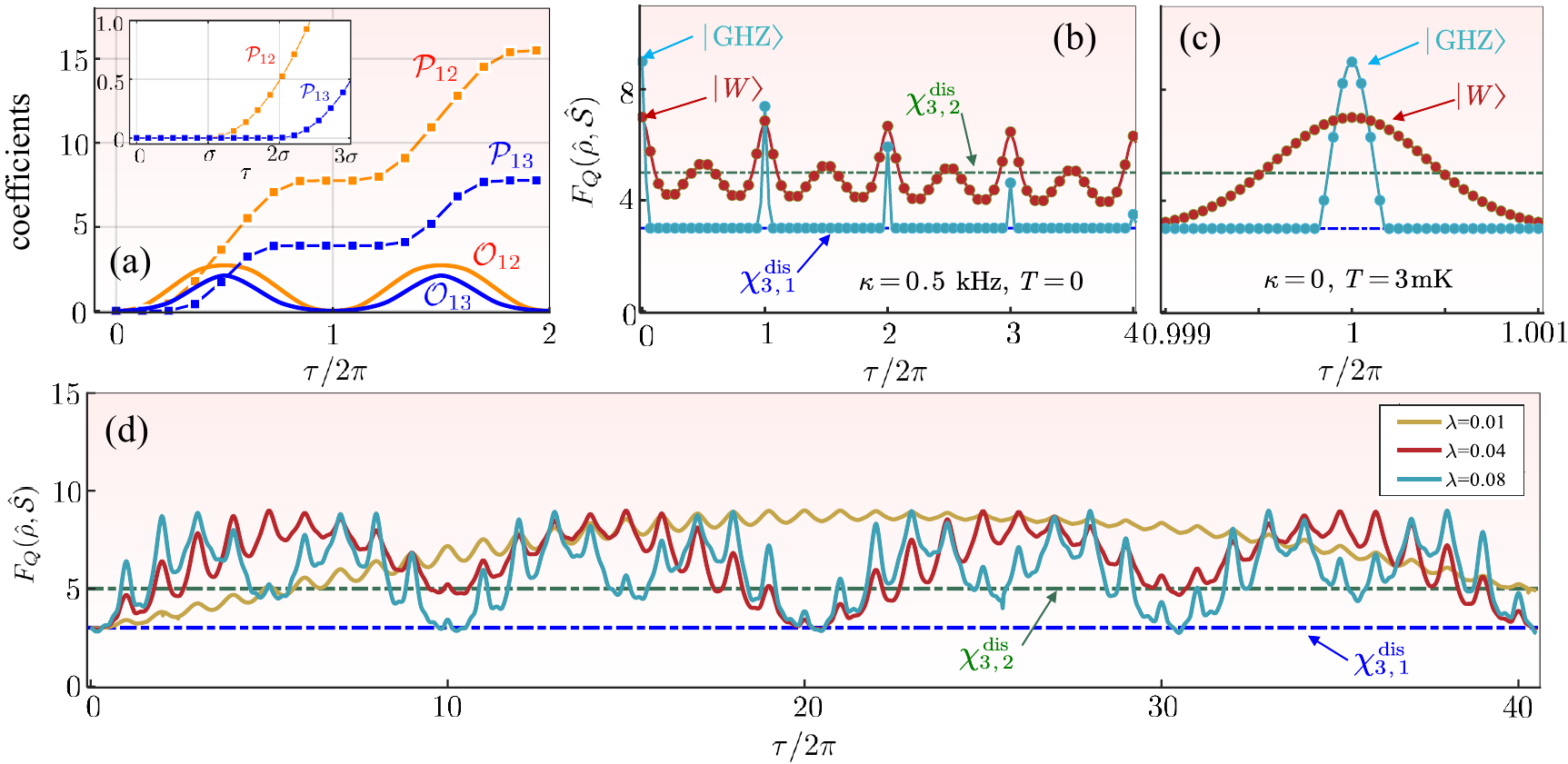}
  \caption{(a) The environment induced mutual phase terms $\mathcal{P}_{kk'}$ (marked lines) and dephasing $\mathcal{O}_{kk'}$ (solid lines) as a function of scaled time $\tau/2\pi=\omega_{1}t/2\pi$ for $N=3$ oscillators with $\lambda=1$. Zoomed figure focuses on the time domain limited to the range of $\tau=0$ to $\tau=3\sigma$. The time evolution of QFI versus dimensionless time $\tau/2\pi$ by considering different initial states $\ket{{\rm GHZ}}$ and $\ket{W}$ with $\kappa=0.5 {{\rm kHz}}$ and $T=0$ in (b) and $\kappa=0$ and $T=3{{\rm mK}}$ in (c). The $F_{Q}(\hat{\rho},\hat{\mathcal{S}})$ versus scaled time $\tau/2\pi$ for $N=3$ oscillators with different parameters $\lambda=0.01,0.04,0.08$ is plotted in (d) with $T=0, \kappa=0$. Here we classify multipartite quantum states into $N$ categories, i.e., separable state, 2-part entanglement state, and 3-part entanglement state. The distance between neighboring LC circuits is chosen as $\sigma=\pi/2(N-1)$ such that the oscillators are only distributed along the strip of half length. Other parameters used here are $\Omega/2\pi= {\rm{15 GHz}}$, $\omega_{1}=50 {\rm {kHz}}$, $g_{1,1}= -10.6 {\rm {kHz}}.$ }\label{figS1}
\end{figure}

In Fig.\,\ref{figS1}(a), we plot the coefficients $\mathcal{P}_{kk'}(t)$ and $\mathcal{O}_{kk'}(t)$ versus scaled time $\tau/2\pi=\omega_{1}t/2\pi$ for $N=3$. It is clear that the dephasing term $\mathcal{O}_{kk'}(t)$ in the early time are stronger than $\mathcal{P}_{kk'}(t)$ resulting a delayed establishment of entanglement distributed among oscillators, even though the terms $\mathcal{P}_{kk'}(t)$ begin to work as soon as $\tau>|k-k'|\sigma$ [see the inset figure in Fig.\,\ref{figS1}(a)]. At the critical instants $t=2n\pi$ with $n$ an arbitrary integer, the dephasing $\mathcal{O}_{kk'}(t)$ are completely suppressed indicating the emergence of full rephasing. To demonstrate how the causal correlation  between $k^{{\rm{th}}}$ and $k'^{{\rm{th}}}$ LC oscillators is guaranteed by  $\mathcal{P}_{kk'}(t)$, it is conducive to rewritten Eq.\,(\ref{S18}) as
\begin{align}
\mathcal{P}_{kk'}(t)=&\lambda\big\{\tau\pi^{2}\mathrm{B}_{2}(\frac{\sigma(k-k')}{2\pi})-\tau\pi^{2}\mathrm{B}_{2}(\frac{\sigma(N+1-k-k')+\pi}{2\pi})
-\frac{\pi^{3}}{3}\mathrm{B}_{3}(\frac{\tau+\sigma(k-k')}{2\pi})-\frac{\pi^{3}}{3}\mathrm{B}_{3}(\frac{\tau-\sigma(k-k')}{2\pi})\nonumber\\
&+\frac{\pi^{3}}{3}\mathrm{B}_{3}(\frac{\tau+\sigma(N+1-k'-k)+\pi}{2\pi})+\frac{\pi^{3}}{3}\mathrm{B}_{3}(\frac{\tau-\sigma(N+1-k'-k)+\pi}{2\pi})\big\},\label{S20}
\end{align}
where $\mathrm{B}_{n}(\bigcdot)$ denotes the $n^{\rm{th}}$ Bernoulli polynomial. Taking parameters $N=3$ and $\sigma=\pi/4$ (i.e., the distance between neighboring resonators is $D=L/4$) as an example, Eq.\,(\ref{S20}) can be simplified in the early time as follows
\begin{align}
\mathcal{P}_{12}(t)=\left\{\begin{array}{lll}0 & & \tau<\sigma\\
\frac{\lambda\pi(\sigma-\tau)^{2}}{4}& & \sigma<\tau<3\sigma\\
-\frac{\lambda\pi(\pi-2\sigma)(\pi-2\tau)}{4}& &3\sigma<\tau<5\sigma\\
\lambda\pi^{2}[\tau-\frac{\pi}{2}-\frac{(\sigma+\tau)^{2}}{4\pi}]& &5\sigma<\tau<7\sigma\\
\frac{1}{2}(\pi-2\sigma)& &7\sigma<\tau<9\sigma\end{array}\right.,
\,\,\mathcal{P}_{13}(t)=\left\{\begin{array}{lll}0 & & \tau<2\sigma\\
\frac{\lambda\pi}{4}(\tau-2\sigma)^{2}& & 2\sigma<\tau<4\sigma\\
-\frac{\lambda\pi}{4}(2\pi^{2}-4\sigma^{2}-4\pi\tau+4\sigma\tau+\tau^{2})& &4\sigma<\tau<6\sigma\\
\frac{\lambda\pi}{2}(\pi-2\sigma)^{2}& &6\sigma<\tau<8\sigma\end{array}\right..\label{S21}
\end{align}
From Eq.\,(\ref{S21}) we can clearly see that the terms $\mathcal{P}_{kk'}(t)$, which are responsible for the establishment of quantum entanglement between $k^{\rm{th}}$ and $k'^{\rm{th}}$ resonators, begin to work if and only if $\tau>|k-k'|\sigma$ as a signature of satisfaction of causality.

It is also note that the dephasing $\mathcal{O}_{kk'}(t)$ is a summation over all of the temperature-dependent factors $\coth(\frac{\beta\hbar\omega_{j}}{2})$. Specifically, these terms can also be expressed analytically in the limit of zero temperature
\begin{align}
\underset{T\rightarrow 0}{\rm{lim}}\mathcal{O}_{kk'}(t)=&\lambda\Re\big[\mathrm{Li}_{3}(e^{i\sigma(k-k')})-\mathrm{Li}_{3}(-e^{i\sigma(N+1-k-k')})]
-\frac{1}{2}\mathrm{Li}_{3}(e^{i[\tau+\sigma(k-k')]})+\frac{1}{2}\mathrm{Li}_{3}(-e^{i[\tau+\sigma(N+1-k'-k)]})\nonumber\\
&-\frac{1}{2}\mathrm{Li}_{3}(e^{i[\tau-\sigma(k-k')]})+\frac{1}{2}\mathrm{Li}_{3}(-e^{i[\tau-\sigma(N+1-k'-k)]})\big].\label{S22}
\end{align}

\section{ DISSIPATIVE DYNAMICS OF  \boldmath$N$ LC RESONATORS}
\setcounter{equation}{22}
\renewcommand\theequation{S\arabic{equation}}
\makeatletter
\renewcommand{\thefigure}{S\@arabic\c@figure}
\makeatother
To demonstrate how the leakage of  photons modifies the generation of multipartite quantum entanglement, we provide detail derivation of  Eq.(4) in the main text by considering a common external dissipation rate $\kappa$ for each of individual resonators.

Here the dissipation rate is required to be small enough so that $\kappa t\ll 1$. Since we only interest in the early dynamical behavior and the dissipation rate of mechanical modes is much less than that of microwave cavity modes, then the damping of mechanical modes can be ignored. Therefore, the master equation of the overall considered open system is
\begin{align}
\dot{\hat{\rho}}_{{\rm tot}}=&i[\hat{\rho}_{{\rm tot}},\hat{H}]+\chi(\hat{\rho}_{{\rm tot}})\nonumber\\
=&i[\hat{\rho}_{{\rm tot}},\hat{H}]+\sum\limits_{k=1}^{N}\frac{\kappa}{2}(2\hat{a}_{k}\hat{\rho}_{{\rm tot}}\hat{a}_{k}^{\dagger}-\hat{a}_{k}^{\dagger}\hat{a}_{k}\hat{\rho}_{{\rm tot}}-\hat{\rho}_{{\rm tot}}\hat{a}_{k}^{\dagger}\hat{a}_{k}),\label{S23}
\end{align}
where the first term in the right-hand side of above master equation is exactly solvable with the time evolution operator $\hat{U}(t)$ given by Eq.(\ref{S5}) and fulfilling $i\dot{\hat{U}}(t)=\hat{H}\hat{U}(t)$. In order to give a perturbative description of Eq.\,(\ref{S20}), we introduce a new density operator $\hat{R}=\hat{U}^{\dagger}\hat{\rho}_{{\rm tot}}\hat{U}$. We could solve the master equation associating $\hat{R}$ formally by performing the first Born approximation \,\cite{PhysRevA.55.3042m} with the result reads
\begin{align}
\hat{R}(t)=\hat{R}_{0}+\int_{0}^{t}\tilde{\chi}(\hat{R}_{0},\tau)d\tau,\label{S24}
\end{align}
where the tilde symbol in $\chi$ obtained by replacing all the operators $\hat{a}_{k}$ in $\chi(\hat{\rho}_{{\rm tot}})$ by $\tilde{\hat{a}}_{k}=\hat{U}^{\dagger}\hat{a}_{k}\hat{U}$. Moreover, the operator $\hat{R}_{0}$ is a constant operator contributing to the free evolution solution $\hat{\rho}_{{\rm tot}}^{0}$ of the non-dissipative system as we have done above. Therefore the density matrix $\hat{\rho}_{{\rm tot}}$ to be solved can be evaluated as
\begin{align}
\hat{\rho}_{{\rm tot}}(t)=&\hat{\rho}_{{\rm tot}}^{0}(t)+\hat{\rho}_{{\rm tot}}^{r}(t)\nonumber\\
=&\hat{\rho}_{{\rm tot}}^{0}(t)+\frac{\kappa}{2}\hat{U}(t)\int_{0}^{t}d\tau\sum\limits_{k=1}^{N}(2\tilde{\hat{a}}_{k}\hat{R}_{0}\tilde{\hat{a}}_{k}^{\dagger}-\tilde{\hat{a}}_{k}^{\dagger}\tilde{\hat{a}}_{k}\hat{R}_{0}-\hat{R}_{0}\tilde{\hat{a}}_{k}^{\dagger}\tilde{\hat{a}}_{k})\hat{U}^{\dagger}(t)\nonumber\\
=&\hat{\rho}_{{\rm tot}}^{0}(t)-\frac{\kappa t}{2}\sum\limits_{k=1}^{N}[\hat{a}_{k}^{\dagger}\hat{a}_{k}\hat{\rho}_{{\rm tot}}^{0}(t)+\hat{\rho}_{{\rm tot}}^{0}(t)\hat{a}_{k}^{\dagger}\hat{a}_{k}]+\kappa\hat{U}(t)\int_{0}^{t}d\tau\sum\limits_{k=1}^{N}\tilde{\hat{a}}_{k}\hat{R}_{0}\tilde{\hat{a}}_{k}^{\dagger}\hat{U}^{\dagger}(t).\label{S25}
\end{align}
Using the Baker-Cambell-Hausdorf expansion, the effect of transformation $\hat{U}(t-\tau)$ on $\hat{a}_{k}$ reads
\begin{align}
\hat{U}(t-\tau)\hat{a}_{k}\hat{U}^{\dagger}(t-\tau)=&\exp(2i\sum\limits_{j=1}^{\infty}\frac{g_{k,j}}{\omega_{j}}[x(t)\sin\frac{\omega_{j}t}{2}-x(\tau)\sin\frac{\omega_{j}\tau}{2}])
\exp(-i\sum\limits_{j=1}^{\infty}[f(\omega_{j}t)-f(\omega_{j}\tau)](\sum_{s\neq k}\frac{2g_{s,j}g_{k,j}(n_{s}+\frac{1}{2})}{\omega_{j}^{2}}))\nonumber\\
&\times\exp(-2i\sum\limits_{j=1}^{\infty}[f(\omega_{j}t)-f(\omega_{j}\tau)](\frac{g_{k,j}}{\omega_{j}})^{2}(1+\hat{a}^{\dagger}_{k}\hat{a}_{k})\hat{a}_{k}),\label{S26}
\end{align}
where we have adopted the modified time evolution operator $\hat{U}(t)$ in Eq.\,(\ref{S8})
\begin{align}
\hat{U}(t)=&\exp(i\sum\limits_{j=1}^{\infty}f(\omega_{j}t)[\sum\limits_{k=1}^{N}\frac{g_{k,j}}{\omega_{j}}(n_{k}+\frac{1}{2})]^{2})\exp(-2i\sum\limits_{j=1}^{\infty}(\sum\limits_{k=1}^{N}\frac{g_{k,j}(n_{k}+\frac{1}{2})}{\omega_{j}})\hat{x}(t)\sin\frac{\omega_{j}t}{2})\exp(-i\hat{H}_{0}t).\label{S27}
\end{align}
Here $\hat{H}_{0}$ is free evolution of the resonators plus the phononic field, and the notations $f(x)=x-\sin(x)$ and $\hat{x}(t)=\hat{b}_{j}^{\dagger}e^{-i\frac{\omega_{j}t}{2}}+\hat{b}_{j}e^{i\frac{\omega_{j}t}{2}}$ have been introduced for conciseness. Plugging Eq.\,(\ref{S26}) to Eq.\,(\ref{S25}), we obtain
\begin{align}
\hat{\rho}_{{\rm tot}}(t)=&\hat{\rho}_{{\rm tot}}^{0}(t)-\frac{\kappa t}{2}\sum\limits_{k=1}^{N}[\hat{a}_{k}^{\dagger}\hat{a}_{k}\hat{\rho}_{{\rm tot}}^{0}(t)+\hat{\rho}_{{\rm tot}}^{0}(t)\hat{a}_{k}^{\dagger}\hat{a}_{k}]\nonumber\\
&+\kappa\int_{0}^{t}d\tau\sum\limits_{k=1}^{N}\bigg\{\exp(-2i\sum\limits_{j=1}^{\infty}[f(\omega_{j}t)-f(\omega_{j}\tau)]\frac{g_{k,j}}{\omega_{j}}N_{j}+2i\sum\limits
_{j=1}^{\infty}\frac{g_{k,j}}{\omega_{j}}[\hat{x}(t)\sin\frac{\omega_{j}t}{2}-\hat{x}(\tau)\sin\frac{\omega_{j}\tau}{2}])\hat{a}_{k}\hat{\rho}_{{\rm tot}}^{0}(t)\hat{a}^{\dagger}_{k}\nonumber\\
&\times \exp(+2i\sum\limits_{j=1}^{\infty}[f(\omega_{j}t)-f(\omega_{j}\tau)]\frac{g_{k,j}}{\omega_{j}}N_{j}-2i\sum\limits
_{j=1}^{\infty}\frac{g_{k,j}}{\omega_{j}}[\hat{x}(t)\sin\frac{\omega_{j}t}{2}-\hat{x}(\tau)\sin\frac{\omega_{j}\tau}{2}])\bigg\}.\label{S28}
\end{align}
After some algebra, we found the reduced density matrix of $N$ LC circuits by tracing out the acoustic degrees of freedom, i.e.,
\begin{align}
{\rm Tr_{m}}[\hat{\rho}_{{\rm tot}}]=&\hat{\rho}(t)+\kappa\int_{0}^{t}d\tau\sum\limits_{k=1}^{N}\bigg\{\exp(-2i\sum\limits_{j=1}^{\infty}[f(\omega_{j}t)-f(\omega_{j}\tau)]\frac{g_{k,j}}{\omega_{j}}N_{j})\hat{a}_{k}\hat{\rho}(t)\hat{a}^{\dagger}_{k}
\exp(2i\sum\limits_{j=1}^{\infty}[f(\omega_{j}t)-f(\omega_{j}\tau)]\frac{g_{k,j}}{\omega_{j}}N_{j})\bigg\}\nonumber\\
&-\frac{\kappa t}{2}\sum\limits_{k=1}^{N}[\hat{a}_{k}^{\dagger}\hat{a}_{k}\hat{\rho}(t)+\hat{\rho}(t)\hat{a}_{k}^{\dagger}\hat{a}_{k}]
,\label{S29}
\end{align}
where $\hat{\rho}(t)={\rm Tr_{m}}[\hat{\rho}^{0}_{{\rm tot}}]$ is exactly the result shown in Eq.\,(\ref{S17}). Therefore, using Eq.\,(\ref{S29}), density matrix elements of $\tilde{\hat{\rho}}(t)\equiv{\rm Tr_{m}}[\hat{\rho}_{{\rm tot}}]$ can be extracted straightforwardly as follows
\begin{align}
\tilde{\rho}_{\mathbf{n},\mathbf{n'}}(t)=&\rho_{\mathbf{n},\mathbf{n'}}(t)
+\kappa\int_{0}^{t}d\tau\sum\limits_{k=1}^{N}\exp(-2i\sum\limits_{j=1}^{\infty}[f(\omega_{j}t)-f(\omega_{j}\tau)]\frac{g_{k,j}}{\omega_{j}}\sum\limits_{s=1}^{N}\frac{g_{s,j}}{\omega_{j}}(n_{s}-n'_{s}))\nonumber\\
&\times\sqrt{(n_{k}+1)(n'_{k}+1)}\rho(t)_{\{n_{1},\cdots,n_{k+1},\cdots,n_{N}\},\{n'_{1},\cdots,n'_{k+1},\cdots,n'_{N}\}}-\frac{\kappa t}{2}\sum\limits_{k=1}^{N}(n_{k}+n'_{k})\rho_{\mathbf{n},\mathbf{n'}}(t)\nonumber\\
=&\rho_{\mathbf{n},\mathbf{n'}}(t)
+\kappa\int_{0}^{t}d\tau\sum\limits_{k=1}^{N}\exp(-2i\sum\limits_{s=1}^{\infty}[\mathcal{P}_{ks}(t)-\mathcal{P}_{ks}(\tau)](n_{s}-n'_{s}))\sqrt{(n_{k}+1)(n'_{k}+1)}\rho_{\mathbf{\tilde{n}}_{k},\mathbf{\tilde{n}'}_{k}}(t)\nonumber\\
&-\frac{\kappa t}{2}\sum\limits_{k=1}^{N}(n_{k}+n'_{k})\rho_{\mathbf{n},\mathbf{n'}}(t).\label{S30}
\end{align}
This is the Eq.(4) in the main text, where the notation $\mathbf{\tilde{n}}_{k}=\{n_{1},\cdots,n_{k-1}, n_{k}+1,n_{k+1},\cdots, n_{N}\}$ has been introduced for simplicity.
\section{ DETECTION AND CLASSIFICATION OF MULTIPARTITE ENTANGLEMENT}
\setcounter{equation}{30}
\renewcommand\theequation{S\arabic{equation}}
\makeatletter
\renewcommand{\thefigure}{S\@arabic\c@figure}
\makeatother
In this section, we show respectively the detailed process of classifying and detecting multipartite entanglement (ME) for discrete and continuous variables, based on quantum fisher information (QFI).  Furthermore, we also adopt criterion from correlation tensors as a comparison, from which we demonstrate that the results of two methods in detecting entanglement are consistent.

\subsection{Multipartite entanglement detection for discrete variables from QFI}
Quantum fisher information (QFI) establish a quantitative connection between entanglement and quantum metrology. It has been widely used to detect the ME for qubits systems.  The QFI associated with operator $\hat{A}$, which we denote $F_{Q}(\hat{\rho},\hat{A})$, expresses how sensitively the state $\hat{\rho}(\theta)=\exp(-i\hat{A}\theta)\hat{\rho}\exp(i\hat{A}\theta)$ changes upon small variations of $\theta$\,\cite{Gessner2017entanglementm}.
It turns out that the highest sensitivities in quantum interferometry are accessible with genuine multiparticle entanglement\,\cite{PhysRevA.85.022321m}. Therefore, QFI provides a powerful and promising tool in accurate classification and detection of various kinds of entangled states.

Given the spectral decomposition for a mixed state $\hat{\rho}=\sum\limits_{k} p_{k}\ket{k}\bra{k}$ with $p_{k}>0$ and $\sum_{k}p_{k}=1$, the QFI can be evaluated as\,\cite{PhysRevA.88.014301m, Naturephys3700m, PhysRevA.85.022321m}
\begin{align}
F_{Q}(\hat{\rho},\mathcal{\hat{S}})=\sum\limits_{p_{k}+p_{l}>0}\frac{2(p_{k}-p_{l})^{2}}{p_{k}+p_{l}}|\bra{k}\mathcal{\hat{S}}\ket{l}|^{2},\label{S31}
\end{align}
where $\mathcal{\hat{S}}=\frac{1}{2}\sum\limits_{l=1}^{N}\mathbf{n}_{l}\cdot\boldsymbol{\sigma}_{l}$ is the linear observable for $N$ qubits system. Here $\boldsymbol{\sigma}_{l}=(\sigma^{x}_{l},\sigma^{y}_{l},\sigma^{z}_{l})$ is a vector containing the Pauli matrices associated with spin $l$ and $\mathbf{n}_{l}$ is a unit vector on the Bloch sphere. The $N$ vectors $\mathbf{n}_{l}$ need to be optimized with $N$ constrants $\mathbf{n}_{l}^{T}\mathbf{n}_{l}=1\,(l=1,2,...,N)$ to maximize $F_{Q}(\hat{\rho},\mathcal{\hat{S}})$. A widely used criterion based on QFI states that the system hosts at least $(m+1)$-particle entanglement if QFI fulfils $F_{Q}(\hat{\rho},\mathcal{\hat{S}})>\chi^{{\rm dis}}_{N,m}=sm^{2}+r^{2}$, where $s=\lfloor\frac{N}{m}\rfloor$ is the largest integer smaller than or equal to $N/m$ and $r=N-sm$\,\cite{PhysRevA.85.022321m}.

We are now in a position to elaborate the prominent effects of phononic environment on ME dynamics. In contrast to the scenario of bipartite case\,\cite{PhysRevLett.129.203604m},  the role of dissipation and environment temperature in ME stands out in its hierarchical structure and high sensitivity. For the former, the increase of dissipation changes gradually the entanglement class of the multipartite states generated at critical instants $t_{p}=2p\pi/\omega_{1}$ from GME to unentangled state in an irreversible way, i.e., a decreasing Schmidt rank [see the numerical results along the line determined by $\tau/2\pi=1$ in Fig.3(b) from the main text]. In this way, adjustable many-body resource is available by engineering system dissipation. We recall that each type of entangled states has its own advantages in performing quantum tasks, and higher-hierarchy entangled states are not necessarily better than lower-hierarchy entangled states. Similar phenomenon of ME stratification occurs in the vicinity of the critical instants $t_{p}=2p\pi/\omega_{1}$  [see the numerical results along the line determined by $\tau/2\pi=1+0.6\varepsilon$ in Fig.3(c) from the main text] by gradually increasing the environment temperature. The ME created at critical instants $t_{p}$ is immune to environment temperature and hence cannot function as ME tunability. For the latter, we point out that the dissipative dynamics of ME is highly sensitive to the specific hierarchy to which the initial quantum state belongs. As shown in  Fig.\,\ref{figS1}(b) and (c), we plot the QFI versus dimensionless time $\tau/2\pi$ by considering initial states belonging to two completely different hierarchies, from which several  observations can be summarized. Firstly, the larger value of QFI for state $\ket{{\rm GHZ}}$ ($\tau=0$) indicates that the three-qubit GHZ state is the most entangled with respect to the ME measure in metrology. Secondly, the entanglement of the states resulting from the time evolution of state $\ket{{\rm GHZ}}$ vanishes completely when dissipation is considered, except for the instants $t_{p}$ at which full rephasing is allowed, while the one for $\ket{W}$ still survives during the considered time scale. This distinct behavior can be explained from the topological viewpoint of ME, i.e., the two in-equivalent tripartite states $\ket{{\rm GHZ}}$ and $\ket{W}$ are respectively 0-resistant and 1-resistant\,\cite{Kauffman_2002m,PhysRevA.98.062335m, PhysRevA.98.062335m, PhysRevD.107.126005m}. Thirdly, the quantity of entanglement at those critical instants $t_{p}$ decreases linearly with $p$, and the one for GHZ state decays more rapidly. Lastly but not least, for the states evolved from GHZ state at time around $t_{p=1}$, the time window during which entanglement appears is more tightly compressed compared to that from the W state.

To better comprehend the fundamental characteristics of the ME dynamics, we here simulate the zero temperature entanglement dynamics with above criterion by assuming the initial state of the system is prepared at $\frac{1}{2^{N/2}}(\ket{0}+\ket{1})^{\otimes N}$.  As shown in Fig.\,\ref{figS1}(d), we plot the dynamics of ME as witnessed by QFI when gradually varying the dimensionless constant $\lambda$ from $\lambda=0.01$ to $\lambda=0.08$.  On the one hand, although the terms $\mathcal{P}_{kk'}$ begin to work as soon as $\tau>\sigma$, the entanglement is established later than expected as a consequence of supernumerary time needed to counteract dephasing process. On the other hand, it is obvious that the larger the constant $\lambda$, the faster the QFI grows, due to the fact that the $\lambda$-dependent $\mathcal{P}_{kk'}$ closely decides the speed or the strength of correlation generation between the $k ^{{\rm th}}$ and $k ^{{\prime\rm th}}$ LC oscillators for $k\neq k'$. More concretely, the curves for $\lambda=0.08$ and $\lambda=0.04$ reach the maximum QFI at $\tau/2\pi=3$ and $\tau/2\pi=5$, respectively, whereas the one for $\lambda=0.01$ reaches its maximum QFI at roughly about $\tau/2\pi=20$. After which the QFI evolves with decreasing peaks in critical instants $\tau=\tau_{n}$ before reaching a minimum peak. The cycle of oscillatory growth and decay for the QFI will continue to repeat indefinitely. In addition, since the dephasing terms $\mathcal{O}_{kk'}$ oscillate periodically in time and completely vanish at $\tau_{n}$ [see Fig.\,\ref{figS1}(a)], a general oscillating behavior for QFI in all of our numerical simulations can be explained, as shown in Fig.\,\ref{figS1}(d).

\subsection{Multipartite entanglement detection for continuous variables from QFI}
In this subsection, we elaborate briefly the method for detecting ME for continuous variables \,\cite{PhysRevA.94.020101m} by comparing QFI $F_{Q}(\hat{\rho},\hat{A}(\mathbf{c}))$ with different bounds $\chi^{{\rm con}}_{N,n}$, where $\hat{A}(\mathbf{c})=\sum_{i=1}^{N}\mathbf{c}_{i}\cdot\hat{\mathbf{A}}_{i}$ is a $N$-body system operator depended by the combined vector $\mathbf{c}=(\mathbf{c}_{1},...,\mathbf{c}_{N})^{T}$. The key step to construct strongest possible criterion is the choice of local operators. In this work, we have adopted $\mathcal{A}=\{\hat{\mathbf{A}}_{1},...,\hat{\mathbf{A}}_{N}\}$ with $\hat{\mathbf{A}}_{i}=[\hat{x}_{i},\hat{p}_{i},\hat{x}^{2}_{i},\hat{p}^{2}_{i},(\hat{x}_{i}\hat{p}_{i}+\hat{p}_{i}\hat{x}_{i})/2]^{T}$, where $\hat{x}_{i}$ and $\hat{p}_{i}$ are the position and momentum operators of $i^{\rm th}$ subsystem, respectively. For instance, the bounds $\chi^{{\rm con}}_{N,n}$ for three-mode systems have the form of \,\cite{PhysRevApplied.18.024065m}
\begin{align}\label{S32}
\chi^{{\rm con}}_{3,1}&=4[{\rm{Var}}(\hat{A}_{1})_{\hat{\rho}_{1}}+{\rm{Var}}(\hat{A}_{2})_{\hat{\rho}_{2}}+{\rm{Var}}(\hat{A}_{3})_{\hat{\rho}_{3}}],\nonumber\\
\chi^{{\rm con}}_{3,2}&=4\underset{i=1,2,3}{{\rm{Max}}}\{{\rm{Var}}(\hat{A}_{i})_{\hat{\rho}_{i}}+{\rm{Var}}(\hat{A}_{j}+\hat{A}_{k})_{\hat{\rho}_{jk}}\},
\end{align}
for $i\neq j\neq k$, where $\hat{A}_{i}=\mathbf{c}_{i}\cdot\hat{\mathbf{A}}_{i}$ is the local operator acting on the reduced state $\hat{\rho}_{i}$ and ${\rm{Var}}(\hat{A})_{\hat{\rho}}=\langle\hat{A}^{2}\rangle_{\hat{\rho}}-\langle\hat{A}\rangle^{2}_{\hat{\rho}}$ denotes the variance.  Then $F_{Q}(\hat{\rho},\hat{A}(\mathbf{c}))>\chi^{{\rm con}}_{3,1}$ predicts that there is entanglement between three modes and $F_{Q}(\hat{\rho},\hat{A}(\mathbf{c}))>\chi^{{\rm con}}_{3,2}$  indicates that the state $\hat{\rho}$ is fully inseparable. The left problem is how to find an optimal vector $\mathbf{c}$ making $F_{Q}(\hat{\rho},\hat{A}(\mathbf{c}))$ go beyond $\chi^{{\rm con}}_{3,n}$ as much as possible. For this purpose, we express the QFI in matrix form $F_{Q}(\hat{\rho},\hat{A}(\mathbf{c}))=\mathbf{c}^{T}Q^{\mathcal{A}}_{\hat{\rho}}\mathbf{c}$ with entries
\begin{align}
\!\!(Q^{\mathcal{A}}_{\hat{\rho}})^{mn}_{ij}=2\!\!\sum\limits_{p_{k}+p_{l}>0}\frac{(p_{k}-p_{l})^{2}}{(p_{k}+p_{l})}\bra{k}\hat{\mathbf{A}}_{i}^{(m)}\ket{l}\bra{l}\hat{\mathbf{A}}_{j}^{(n)}\ket{k},\label{S33}
\end{align}
where $\hat{\mathbf{A}}_{i}^{(m)}$ is the $m^{\rm th}$ local operator in $\hat{\mathbf{A}}_{i}$,  $p_{k}$ and $\ket{k}$ constitute the spectral decomposition $\hat{\rho}=\sum\limits_{k} p_{k}\ket{k}\bra{k}$.  The indices $i$ and $j$ in $(Q^{\mathcal{A}}_{\hat{\rho}})^{mn}_{ij}$ represent different Hilbert spaces for resonators $i$ and $j$, respectively. Following the same way, the bounds $\chi^{{\rm con}}_{N,n}$ can also be converted into matrix form with elements $(\Gamma^{\mathcal{A}}_{\chi^{{\rm con}}_{N,n}(\hat{\rho})})^{mn}_{ij}=4\,{\rm{Cov}}(\hat{\mathbf{A}}_{i}^{(m)},\hat{\mathbf{A}}_{j}^{(n)})_{\hat{\rho}}$, where ${\rm{Cov}}(\hat{A},\hat{B})_{\hat{\rho}}=\langle \hat{A}\hat{B}+\hat{B}\hat{A}\rangle_{\hat{\rho}}/2-\langle\hat{A}\rangle_{\hat{\rho}}\langle \hat{B}\rangle_{\hat{\rho}}$ is the covariance matrix. For example, only the block-diagonal elements ($i = j$) are allowed to be non-zero for matrix $(\Gamma^{\mathcal{A}}_{\chi^{{\rm con}}_{3,1}(\hat{\rho})})^{mn}_{ij}$.  For a more complicated case of  $4[{\rm{Var}}(\hat{A}_{1})_{\hat{\rho}_{1}}+{\rm{Var}}(\hat{A}_{2}+\hat{A}_{3})_{\hat{\rho}_{23}}]$,  except for the diagonal block elements, the off-diagonal block elements ($i = 2, j=3$; and $i=3, j=2$) in matrix $(\Gamma^{\mathcal{A}}_{\chi^{{\rm con}}_{3,2}(\hat{\rho}_{1,23})})^{mn}_{ij}$ can also be non-empty. After that we could maximize
\begin{align}
W_{N,n}(\hat{\rho},\hat{A}(\mathbf{c}))=F_{Q}(\hat{\rho},\hat{A}(\mathbf{c}))-\chi^{{\rm con}}_{N,n} \label{S34}
\end{align}
by variation of $\mathbf{c}$ to obtain an optimized entanglement witness for the state $\hat{\rho}$. The optimal vector $\mathbf{c}_{{\rm{opt}}}$ is the eigenvector corresponds to the largest eigenvalue of $W_{N,n}(\hat{\rho},\hat{A}(\mathbf{c}))$, and the optimal witness system operator $\hat{A}_{{\rm{opt}}}$ is given by $\mathbf{c}_{{\rm{opt}}}^{T}\mathcal{A}$.

\begin{figure}
  \centering
  \includegraphics[width=17cm]{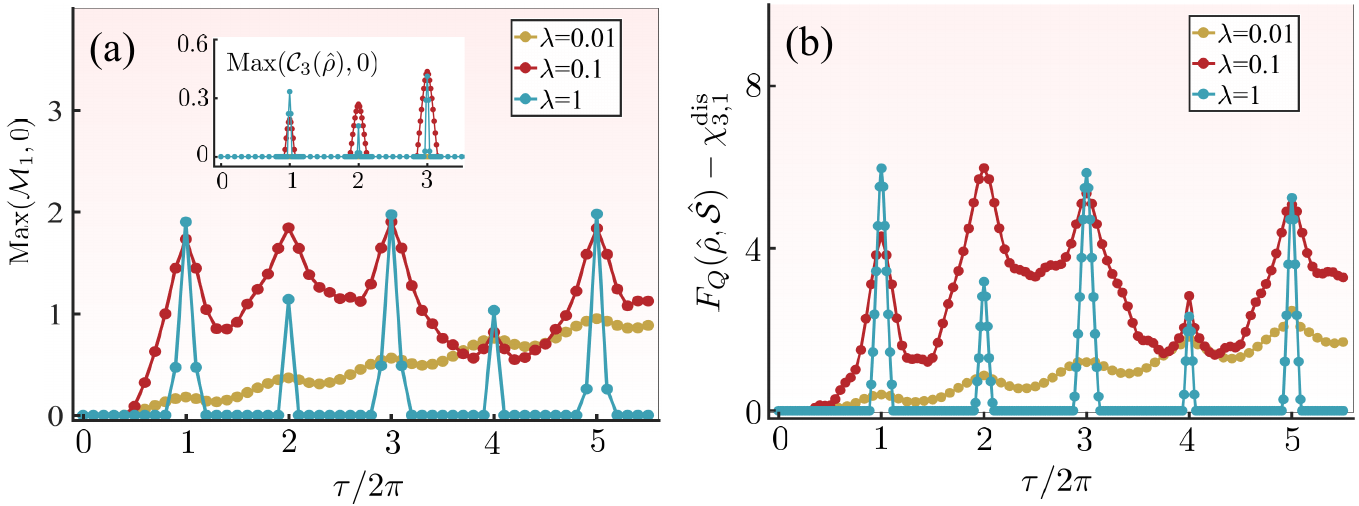}
  \caption{ Evolution of (a) ${\rm{Max}}(\mathcal{M}_{1},0)$  and (b) $F_{Q}(\hat{\rho},\hat{\mathcal{S}})-\chi^{{\rm dis}}_{3,1}$ as a function of scaled time $\tau/2\pi=\omega_{1}t/2\pi$ by altering the parameters $\lambda$. Zoomed figure in (a) depicts the quantity ${\rm{Max}}(\mathcal{C}_{3}(\hat{\rho}),0)$ as a function of scaled time $\tau/2\pi$ . The distance between neighboring LC circuits is chosen as $\sigma=\pi/4$ such that the oscillators are only distributed along the strip of half length. Other parameters used here are $T=0$, $\Omega/2\pi= {\rm{15 GHz}}$, $\omega_{1}=50 {\rm {kHz}}$, $g_{1,1}= -10.6 {\rm {kHz}}.$ }\label{figS2}
\end{figure}

\subsection{Multipartite entanglement detection for discrete variables from correlation tensors}
In this subsection, we apply the alternative entanglement witness, i.e.,  the so-called correlation tensors, to identify the non-fully separated state and also the genuinely entangle state. Before which we briefly introduce several notations and definitions.

Considering a $N$-partite state $\hat{\rho}$ in the Hilbert space $\mathcal{H}^{d}_{1}\otimes\mathcal{H}^{d}_{2}\otimes\cdots\otimes\mathcal{H}^{d}_{N}$ sharing the same local dimension $d$, we select the generators $\hat{\lambda}_{i}$ $(i=1,2,\cdots,d^{2}-1)$ from ${\rm{SU}}(d)$ as the operators to carry out the correlated measurement among the $N$ LC oscillators. The generators $\hat{\lambda}_{i}$ $(i=1,2,3)$ of two level systems are simply the Pauli spin matrices, and the ones related with the general $d$-level systems are the well known Gell-Mann operators. All of these generators combined with the correlation tensors give the Bloch representation or multipartite Fano form of density operators. Arbitrary quantum state $\hat{\rho}$ can be expanded with all of the possible combination of $\hat{\lambda}_{i}$, which has the general form of\,\cite{PhysRevA.84.062306m, Eur.Phys.J.Plusm, PhysRevA.96.052314m}
\begin{align}
\hat{\rho}=&\frac{1}{d^{n}}\hat{I}\otimes\cdots\otimes\hat{I}+\frac{1}{2d^{N-1}}(\sum\limits_{i_{1}=1}^{d^{2}-1}t_{i_{1}}\hat{\lambda}_{i_{1}}\otimes\cdots\otimes\hat{I}
+\cdots\sum\limits_{i_{N}=1}^{d^{2}-1}t_{i_{N}}\hat{I}\otimes\cdots\otimes\hat{\lambda}_{i_{N}})\nonumber\\
&+\cdots+\cdots\frac{1}{2^{N}}\sum\limits_{i_{1}\cdots i_{N}}^{d^{2}-1}t_{i_{1}\cdots i_{N}}\hat{\lambda}_{i_{1}}\otimes\hat{\lambda}_{i_{2}}\otimes\cdots\otimes \hat{\lambda}_{i_{N}}.\label{S35}
\end{align}
Here $\hat{I}$ denotes the identity operator and the coefficients before tensor products are elements of  the correlation tensors
\begin{align}
&t_{i_{1}}={\rm Tr}[\rho\hat{\lambda}_{i_{1}}\otimes\hat{I}\otimes\cdots\otimes\hat{I}],\label{S36}\\
&t_{i_{j_{1}}\cdots i_{j_{k}}}={\rm Tr}[\rho\hat{I}\otimes\cdots\otimes\hat{\lambda}_{i_{j_{1}}}\otimes\cdots\otimes\hat{\lambda}_{i_{j_{k}}}\otimes\hat{I}],\label{S37}\\
&t_{i_{1}\cdots i_{N}}={\rm Tr}[\rho\hat{\lambda}_{i_{1}}\otimes\hat{\lambda}_{i_{2}}\otimes\cdots\otimes \hat{\lambda}_{i_{N}}],\label{S38}
\end{align}
where the subscripts $j_{s}$ of $t_{i_{j_{1}}\cdots i_{j_{k}}}$ represent the position of $\hat{\lambda}_{i_{j_{s}}}$ in the tensor products, and $\mathcal{T}_{i_{1}\cdots i_{N}}$ called full correlation tensor with entries $t_{i_{1}\cdots i_{N}}$. To detect the ME using the criteria based on various matrix norms, it is necessary to convert the tensors into matrix forms by letting certain indices joining together to be row indices (underlined) and the remainder be the column indices (non-underlined). For instance, we construct such a matrix $\mathcal{T}_{\underline{j_{1}}j_{2}\cdots j_{N}}$ with the entires
\begin{align}
\mathcal{T}_{i_{j_{1}},(d^{2}-1)^{N-2}(i_{j_{2}}-1)+\cdots+(d^{2}-1)(i_{j_{N}-1}-1)+i_{j_{N}}}=t_{i_{1}\cdots i_{N}}.\label{S39}
\end{align}
With these definitions, we can now readily illustrate the criteria.  In the following, we study ME of a tripartite qubits system, i.e., $N=3$ and $d=2$,  for conciseness.  It has been shown that any matricization of the full correlation tensor of fully separable tripartite state must fulfill the following inequality\cite{PhysRevA.84.062306m}
\begin{align}
\left|\left|\mathcal{T}_{\underline{i_{1} \cdots i_{k}}i_{k+1}\cdots i_{3}}\right|\right|_{{\rm{tr}}}\le \prod_{j=1}^{3}\sqrt{\frac{2(d-1)}{d}},\,k=1,...,2,\label{S40}
\end{align}
where $||\bigcdot||_{{\rm{tr}}}$ denotes the trace norm. Furthermore, a tripartite state $\hat{\rho}$ is genuinely entangled if the follow inequalities\cite{PhysRevA.96.052314m}
\begin{align}
M_{k}(\hat{\rho})>\mathcal{K}_{k}=\frac{2\sqrt{2}}{3}(2\sqrt{k}+1)\frac{d-1}{d}\sqrt{\frac{d+1}{d}} \label{S41}
\end{align}
are satisfied for any $k=1,2,3$, where $M_{q}(\hat{\rho})$ denotes the average matricization norm
\begin{align}
M_{k}(\hat{\rho})=\frac{1}{3}(||\mathcal{T}_{\underline{1}23}||_{k}+||\mathcal{T}_{\underline{2}13}||_{k}+||\mathcal{T}_{\underline{3}12}||_{k}).\label{S42}
\end{align}
and $||\bigcdot||_{k}$ denotes the Ky Fan $k$ norms. By defining quantities
\begin{align}
\mathcal{M}_{1}\equiv{\rm{Min}}(||\mathcal{T}_{\underline{1}23}||_{{\rm{tr}}},||\mathcal{T}_{\underline{2}13}||_{{\rm{tr}}},||\mathcal{T}_{\underline{3}12}||_{{\rm{tr}}})-1 \label{eqS43}\\
\mathcal{C}_{k}(\hat{\rho})\equiv M_{k}(\hat{\rho})-\frac{\sqrt{3}}{3}(2\sqrt{k}+1), \label{eqS44}
\end{align}
then $\mathcal{M}_{1}>0$ and $\mathcal{C}_{k}(\hat{\rho})>0$ could be used to estimate whether or not a given tripartite state is entangled or genuinely entangled\,\cite{PhysRevA.96.052314m}, respectively.

The application of criteria based on  correlation tensors is shown in Fig.\,\ref{figS2}(a) by calculating the time evolution of ${\rm{Max}}(\mathcal{C}_{3}(\hat{\rho}),0)$ where only 3 LC circuits are considered for illustration purpose.  Furthermore, the inset in  Fig.\,\ref{figS2}(a) depicts the quantity ${\rm{Max}}(\mathcal{C}_{3}(\hat{\rho}),0)$ as a function of scaled time $\tau$.  As shown in Fig.\,\ref{figS2}(b) ,  QFI has also been  utilized to detect the ME by determinating the dynamical behavior of $F_{Q}(\hat{\rho},\hat{\mathcal{S}})-\chi^{{\rm dis}}_{3,1}$ according to Eq.(\ref{S31}) as a comparison.  It is clear that the multipartite entanglement dynamics revealed by these two methods fits very well in detecting whether or not the given state is separable.  However, the ability of criteria based on  correlation tensors to detect genuine multipartite entanglement (see inset in Fig.\,\ref{figS2}(a)) is shown to be weaker than that of QFI method (see Fig.2(b) in the main text).

\section{PREPARATION OF MULTIPARTITE CAT STATES IN A LINEAR CHAIN OF \boldmath$N$ LC OSCILLATORS}
\setcounter{equation}{44}
\renewcommand\theequation{S\arabic{equation}}
\makeatletter
\renewcommand{\thefigure}{S\@arabic\c@figure}
\makeatother
In this section, we primarily present an efficient approach of generating multipartite cat states which are known as an elegant demonstration of Schrödinger’s famous cat paradox. In order to visualize the nonclassical quantum states of the field inside the $N$ oscillators, we apply a full quantum state tomography by measuring the joint Wigner function. Below, we illustrate how multi-mode cat states are generated at critical instants by searching for optimal parameters, and provide the detailed derivations of calculating joint Wigner function.

\subsection{Efficient scheme of generating \boldmath$N$-partite cat states}
In this subsection,  we predict the possibility of generating multipartite cat states when the initial state of system + environment is prepared at $\ket{\Psi(0)}\bra{\Psi(0)}\otimes \hat{\rho}_{T}$, where $\ket{\Psi(0)}=\ket{\alpha_{k}}^{\otimes N}$ is the product state of coherent states for $N$ LC resonators and $ \hat{\rho}_{T}$ is a thermal state for the strip with temperature $T$. Here, we denote the corresponding amplitude of coherent state for $k^{{\rm th}}$ resonator as $\alpha_{k}$.  By absorbing the linear terms that are proportional to $\hat{b}^{\dagger}_{j}+\hat{b}_{j}$ into the definition of the equilibrium positions\,\cite{PhysRevLett.93.266403m, PhysRevA.72.041405m} and neglecting the free evolution term $e^{-i\hat{H}_{0}t}$,  the evolution operator can have a compact form:
\begin{align}
\hat{U}(t)&=\exp[i\sum\limits_{j=1}^{\infty}f(\omega_{j}t)(\sum\limits_{k=1}^{N}\frac{g_{k,j}}{\omega_{j}}\hat{n}_{k})^{2}] \exp[-2i\sum\limits_{j=1}^{\infty}(\sum\limits_{k=1}^{N}\frac{g_{k,j}}{\omega_{j}}\hat{n}_{k})\hat{x}(t)\sin(\frac{\omega_{j}t}{2})] \,\nonumber\\
&=\exp[i\sum_{j=1}^{\infty}(\sum_{k=1}^{N}\frac{g_{k,j}\hat{n}_{k}}{\omega_{j}})^{2}[\omega_{j}t-\sin(\omega_{j}t)]]\exp[-\sum_{k=1}^{N}\sum_{j=1}^{\infty}\frac{g_{k,j}\hat{n}_{k}}{\omega_{j}}[\hat{b}_{j}^{\dagger}(1-e^{-i\omega_{j}t})-\hat{b}_{j}(1-e^{i\omega_{j}t})]].\label{S45}
\end{align}
It is  easy to see that the resonators and the strip are disentangled  at critical instants $t_{p}=2p\pi/\omega_{1}$.  As an example, we first consider $N=3$ oscillators that are prepared initially in $\ket{\alpha_{1}}_{c_{1}}\otimes\ket{\alpha_{2}}_{c_{2}}\otimes \ket{\alpha_{3}}_{c_{3}}$. The evolved tripartite state, after tracing out the phononic modes, at time $t_{p}$ is
\begin{align}
\ket{\psi(t=t_{p})}&=\sum_{n=0}^{\infty}\sum_{m=0}^{\infty}\sum_{s=0}^{\infty}\frac{\alpha_{1}^{n}\alpha_{2}^{m}\alpha_{3}^{s}\exp[i\sum\limits_{j=1}^{\infty}\omega_{j}t(\frac{g_{1,j}}{\omega_{j}}n+\frac{g_{2,j}}{\omega_{j}}m+\frac{g_{3,j}}{\omega_{j}}s)^{2}]}{\sqrt{n!m!s!}}|n\rangle_{c_{1}}\otimes|m\rangle_{c_{2}}\otimes|s\rangle_{c_{3}}\,\nonumber\\
&=\sum_{n=0}^{\infty}\frac{\alpha_{1}^{n}}{\sqrt{n!}}\exp[i\sum_{j=1}^{\infty}\frac{g_{1,j}^{2}}{\omega_{j}}n^{2}t]|n\rangle_{c_{1}}\otimes|\xi_{n}\rangle_{c_{2},c_{3}},\label{S46}
\end{align}
where we have defined
\begin{align}
|\xi_{n}\rangle_{c_{2},c_{3}}&=\sum_{m=0}^{\infty}\sum_{s=0}^{\infty}\frac{\alpha_{2}^{m}\alpha_{3}^{s}}{\sqrt{m!s!}}\exp[i\sum_{j=1}^{\infty}\omega_{j}t(\frac{g_{2,j}}{\omega_{j}}m+\frac{g_{3,j}}{\omega_{j}}s)^{2}]
\exp[i\sum_{j=1}^{\infty}2\omega_{j}t(\frac{g_{1,j}}{\omega_{j}}n)(\frac{g_{2,j}}{\omega_{j}}m+\frac{g_{3,j}}{\omega_{j}}s)]|m\rangle_{c_{2}}\otimes|s\rangle_{c_{3}}\,\nonumber\\
&=\sum\limits_{m=0}^{\infty}\frac{\exp[i\sum\limits_{j=1}^{\infty}\omega_{1}t\frac{g_{2,j}^{2}}{\omega_{1}\omega_{j}}m^{2}]\left(\exp[i\sum\limits_{j=1}^{\infty}2\omega_{1}t\frac{g_{1,j}g_{2,j}}{\omega_{1}\omega_{j}}n]\alpha_{2}\right)^{m}}{\sqrt{m!}}|m\rangle_{c_{2}}\otimes|\xi_{m,n}\rangle_{c_{3}},\label{S47}
\end{align}
for simplicity with
\begin{align}
|\xi_{m,n}\rangle_{c_{3}}=\sum\limits_{s=0}^{\infty}\frac{\exp[i\sum\limits_{j=1}^{\infty}\omega_{1}t\frac{g_{3,j}^{2}}{\omega_{1}\omega_{j}}s^{2}](\alpha_{3}\exp[i\sum\limits_{j=1}^{\infty}2\omega_{1}t(\frac{g_{1,j}g_{3,j}}{\omega_{1}\omega_{j}}n+\frac{g_{2,j}g_{3,j}}{\omega_{1}\omega_{j}}m)])^{s}}{\sqrt{s!}}|s\rangle_{c_{3}}.\label{S48}
\end{align}
By introducing the notations $J_{k_{1},k_{2}}\equiv\sum\limits_{j=1}^{\infty}\frac{g_{k_{1},j}g_{k_{2},j}}{\omega_{1}\omega_{j}}$, the state $\ket{\psi(t=t_{p})}$ can be rewritten as
\begin{small}
\begin{align}
|\psi(t_{p})\rangle=\sum_{n=0}^{\infty}\frac{\alpha_{1}^{n}}{\sqrt{n!}}e^{i2p\pi J_{11}n^{2}}|n\rangle_{c_{1}}\otimes\left\{ \sum_{m=0}^{\infty}\frac{e^{i2p\pi J_{22}m^{2}}\left(\alpha_{2}e^{i4p\pi J_{12}n}\right)^{m}}{\sqrt{m!}}|m\rangle_{c_{2}}\otimes\left[\sum_{s=0}^{\infty}\frac{e^{i2p\pi J_{33}s^{2}}(\alpha_{3}e^{i4p\pi(J_{13}n+J_{23}m)})^{s}}{\sqrt{s!}}|s\rangle_{c_{3}}\right]\right\}.\label{S49}
\end{align}
\end{small}
We proceed by setting $J_{mn}=\frac{1}{4p}+N_{mn}$ with $N_{mn}$ an arbitrary integer.  After that, we obtain
\begin{align}
|\psi(t_{p})\rangle&=\sum_{n=0}^{\infty}\frac{\alpha_{1}^{n}}{\sqrt{n!}}e^{i\frac{\pi}{2}n^{2}}|n\rangle_{c_{1}}\otimes\left\{ \sum_{m=0}^{\infty}\frac{e^{i\frac{\pi}{2}m^{2}}\left(\alpha_{2}e^{in\pi}\right)^{m}}{\sqrt{m!}}|m\rangle_{c_{2}}\otimes\left[\sum_{s=0}^{\infty}\frac{e^{i\frac{\pi}{2}s^{2}}\left(\alpha_{3}e^{i(n+m)\pi}\right)^{s}}{\sqrt{s!}}|s\rangle_{c_{3}}\right]\right\}\,\nonumber\\ &=\frac{1}{\sqrt{2}}(e^{i\pi/4}|\alpha_{n}\rangle_{c_{n}}^{\otimes N=3}+e^{-i\pi/4}|-\alpha_{n}\rangle_{c_{n}}^{\otimes N=3}).\label{S50}
\end{align}
Up to now, we have successfully prepared a three-mode cat state when all of the equalities $J_{mn}=\frac{1}{4p}+N_{mn}$ are satisfied. To observe this cat state numerically, we optimize $\sigma$ and $\lambda$ by minimizing $\varepsilon=\sum\limits_{m,n}|e^{i2p\pi J_{mn}}-i|$ with the simulations are presented in the main text. It is worth remarking that the generation approach of tripartite cat state can be extended to more oscillators in a similar manner.

\begin{figure}
  \centering
  \includegraphics[width=17cm]{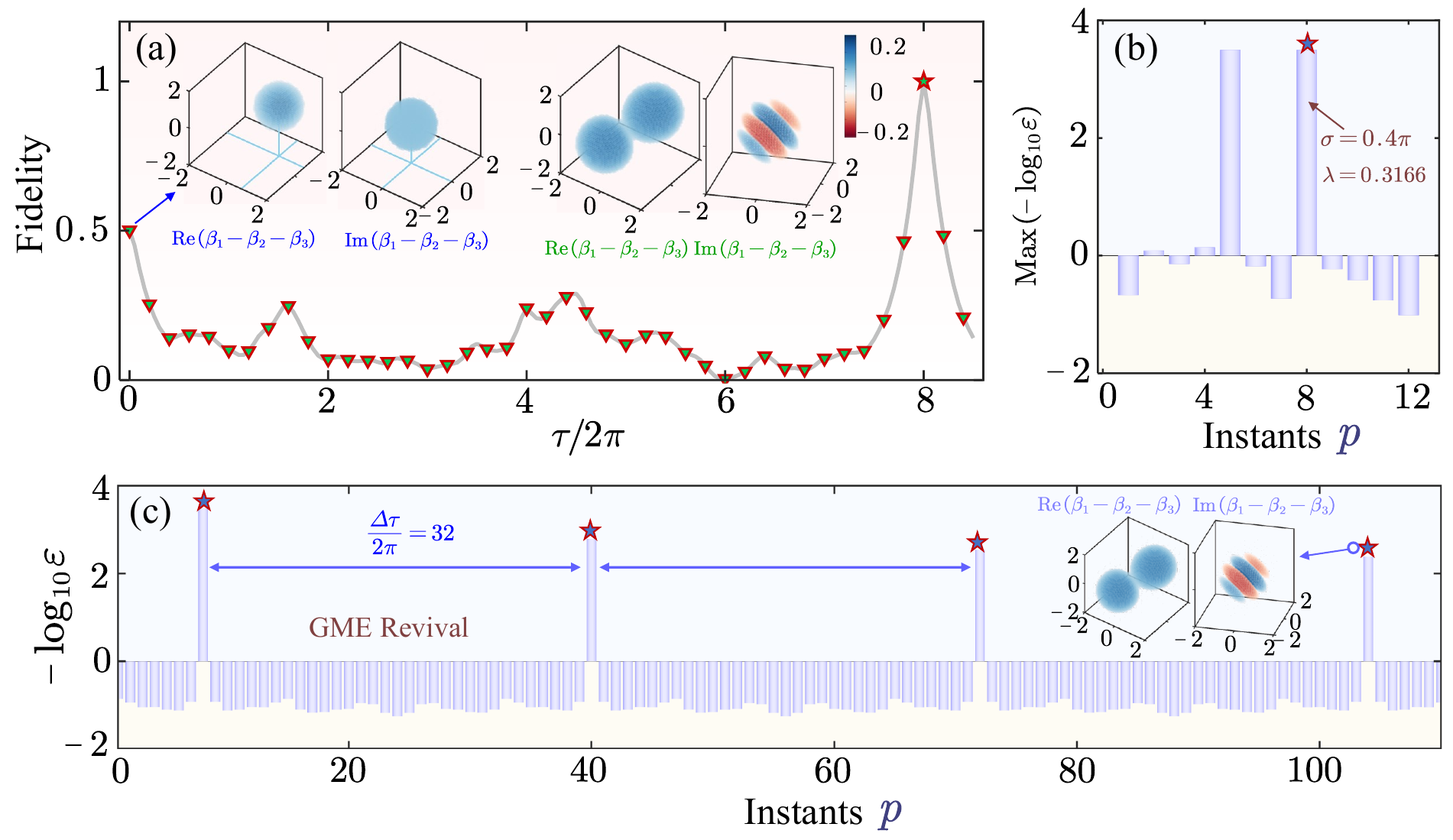}
  \caption{ (a) The fidelity between the states $\ket{\psi_{{\rm cat}}}\bra{\psi_{{\rm cat}}}$ and $\hat{\rho}(t)$ versus scaled time $\tau/2\pi$ with $\sigma=0.4 \pi$ and $\lambda=0.3166$. The insets are the 3D plane-cuts of the measured 6D joint Wigner function for states at $\tau/2\pi=0$ and $\tau/2\pi=8$, respectively. (b) The bar graph of maximal efficiency of generating $\ket{\psi_{{\rm cat}}}$ by scanning the instants $t_{p}$ for $p=0,\cdots,13$. (d) The bar graph of efficiency $-\log_{10}\varepsilon$ as a function of instants $t_{p}$ by fixing the optimal parameters $\sigma=0.4 \pi,\lambda=0.3166$. Other parameters are $N=3, T=0,\alpha=0.7$.  }\label{figS3}
\end{figure}

In Fig.\,\ref{figS3}(a), we simulate the dynamics of fidelity $\mathcal{F}(\hat{\rho}_{{\rm cat}},\hat{\rho}(t))$ between the states $\hat{\rho}_{{\rm cat}}=\ketbra{\psi_{{\rm cat}}}$ and $\hat{\rho}(t)$ for $N=3$, where the fidelity is defined as $\mathcal{F}(\hat{\rho}_{0},\hat{\rho})\equiv{\rm{Tr}}\sqrt{\sqrt{\hat{\rho}}\hat{\rho}_{0}\sqrt{\hat{\rho}}}$. The dimensionless parameter $\lambda$ and scaled distance $\sigma$ are optimized by minimizing the error function $\varepsilon\equiv\sum_{m,n}|e^{2pi\pi J_{mn}}-i|$ as shown in Fig.\,\ref{figS3}(b), from which we find a minimal error of the order $10^{-4}$ at instants $t_{p=5}$ and $t_{p=8}$. The simulation result for the formal has been presented in the main text. Here, we focus on the one for the latter.  We find that such a extremely low conditional error enables us to observe the deterministic generation of tripartite Schrödinger cat state with a fidelity lager than $99.99\%$ [see pentagram-marked data in Fig.\,\ref{figS3}(a)]. The large instant $t_{p=8}$ in this scenario would invalidate the approximate time evolution solution given in Eq.\,(\ref{S30}), thus we do not present the effect of dissipation $\kappa$ on fidelity at $t_{p=8}$ or at revival instants $t_{p=8n}$ for $n=1,2,\cdots$.  The 3D plane-cuts of the measured 6D joint Wigner function for the generated cat state are plotted in the insets of Fig.\,\ref{figS3}(a), featuring prominent non-classical characteristics by negative valued $W_{J}(\{\beta\})$ in the fringes from space $\Im(\beta_{1}-\beta_{2}-\beta_{3})$. More interestingly, our proposed system also allows the remarkable revival of three-mode cat state that are verified by the periodic high peaks in efficiency function $-\log_{10}\varepsilon$. This cycle, for instance, would be $\Delta\tau/2\pi=32$ as shown in Fig.\,\ref{figS3}(c), when fixing optimal parameters obtained in Fig.\,\ref{figS3}(b) at instant $t_{p=8}$.

\subsection{The efficient method of calculating joint Wigner function}
A full quantum state tomography of the $N$ resonator system can be realized by measuring the joint Wigner function $W_{J}(\{\beta\})$\,\cite{science.aaf2941m,sciadv.abn1778m}
 \begin{align}
W_{J}(\{\beta\})=(\frac{2}{\pi})^{N}{\rm{Tr}}[\hat{\rho}\prod_{k=1}^{N}\hat{D}_{k}(\beta_{k})\hat{P}_{k}\hat{D}_{k}^{\dagger}(\beta_{k})],\label{S51}
\end{align}
with $\{\beta\}=\beta_{1},\cdots,\beta_{N}$ being the complex parameters defining the coordinates in the joint phase space.  We proceed by writing down $\hat{\rho}(t) $ in the tensor Fock state basis, truncated to a maximum photon number $N_{{\rm{cut}}}$. Taking 2-partite states for illustration purpose, we have
\begin{align}
\hat{\rho}(t)=\sum\limits_{i,j,m,n=1}^{N_{{\rm{cut}}}}\rho_{mn,ij}(t)\ket{mn}\bra{ij}.\label{S52}
\end{align}
Plugging Eq.\,(\ref{S52}) into Eq.\,(\ref{S51}), we find
\begin{align}
W_{J}(\beta_{1},\beta_{2})&=\frac{4}{\pi^{2}}{\rm{Tr}}[\rho\hat{D}_{1}(\beta_{1})\hat{P}_{1}\hat{D}_{1}^{\dagger}(\beta_{1})\hat{D}_{2}(\beta_{2})\hat{P}_{2}\hat{D}_{2}^{\dagger}(\beta_{2})]\nonumber\\
&=\frac{4}{\pi^{2}}\sum\limits_{i,j,m,n=1}^{N_{{\rm{cut}}}}\rho_{mn,ij}(t)\bra{i}\hat{D}_{1}(\beta_{1})e^{i\pi\hat{a}_{1}^{\dagger}\hat{a}_{1}}\hat{D}_{1}^{\dagger}(\beta_{1})\ket{m}\bra{j}\hat{D}_{2}(\beta_{2})e^{i\pi\hat{a}_{2}^{\dagger}\hat{a}_{2}}\hat{D}_{2}^{\dagger}(\beta_{2})\ket{n}\nonumber\\
&=\frac{4}{\pi^{2}}\sum\limits_{i,j,m,n=1}^{N_{{\rm{cut}}}}\rho_{mn,ij}(t)\bra{i}\hat{D}_{1}(2\beta_{1})\ket{m}\bra{j}\hat{D}_{2}(2\beta_{2})\ket{n}(-1)^{m+n}
.\label{S53}
\end{align}
Similarly,  joint Wigner function for 3-partite states can be obtained by following the same procedures, i.e.,
\begin{align}
W_{J}(\beta_{1},\beta_{2},\beta_{3})=(\frac{2}{\pi})^{3}\sum\limits_{mnp;ijk=1}^{N_{{\rm{cut}}}}\rho_{mnp,ijk}(t)\bra{i}\hat{D}_{1}(2\beta_{1})\ket{m}\bra{j}\hat{D}_{2}(2\beta_{2})\ket{n}\bra{k}\hat{D}_{3}(2\beta_{3})\ket{p}(-1)^{m+n+p}
,\label{S54}
\end{align}
where the matrix elements of displacement operator are give by
\begin{align}
\bra{n}\hat{D}(\alpha)\ket{m}=(\frac{n!}{m!})^{\frac{1}{2}}\exp(-\frac{1}{2}|\alpha|^{2})(-\alpha^{*})^{m-n}L_{n}^{m-n}(|\alpha|^{2})\label{S55}
\end{align}
for $m\ge n$, and $\bra{n}\hat{D}(\alpha)\ket{m}=\bra{m}\hat{D}(-\alpha)\ket{n}^{*}$ for $m<n$. Here $L_{n}^{m-n}(x)$ is a generalized Laguerre polynomial.

\section{UNIQUE PROPERTIES OF MULTIPARTITE ENTANGLED STATES BEYOND BIPARTITE CASE}
\setcounter{equation}{55}
\renewcommand\theequation{S\arabic{equation}}
\makeatletter
\renewcommand{\thefigure}{S\@arabic\c@figure}
\makeatother
In this section, we primarily present the unique properties of multipartite entangled states in terms of their elegant entanglement classification, intricate geometric structure, non-unique quantification and indispensable application for quantum tasks, when comparing with that from bipartite entangled states. Below, we will stress in detail the essential difference between the multipartite entanglement including GME and the bipartite entanglement one by one.

\begin{figure*}
	\includegraphics[width=16cm]{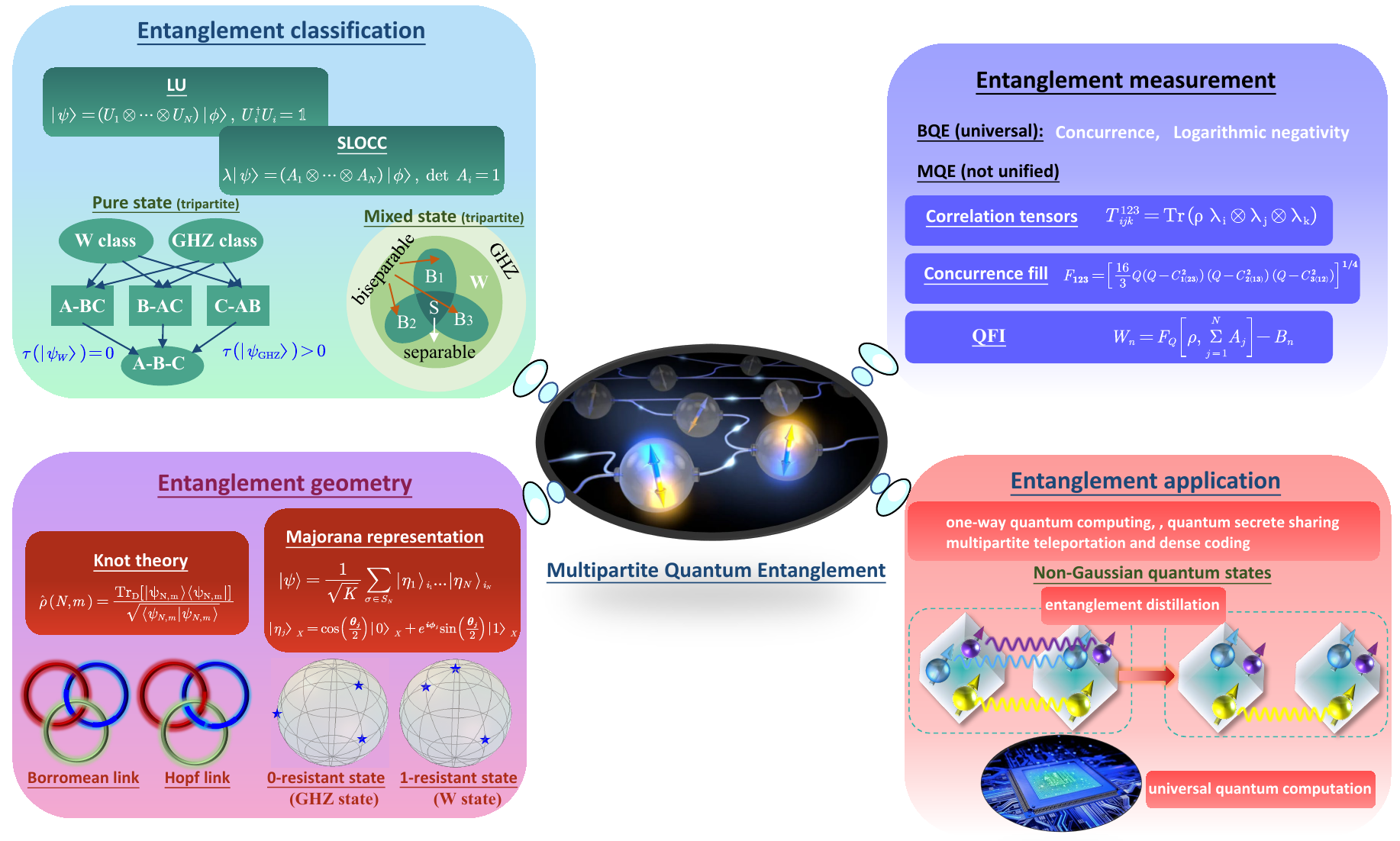}
	\caption{The illustration of unique properties for the multipartite entanglement in the aspects of entanglement classification (background-color: green), entanglement geometry (background-color: mauve), entanglement measurement (background-color: slate blue), and entanglement application (background-color: pink).}
	\label{figS4}
\end{figure*}

Firstly, in terms of entanglement classification, ME exhibits more complex and ingenious entanglement structures compared to BE, even for the simplest case of tripartite case that are included in our work.  In order to determine whether two multipartite entangled states are equivalently useful for a given quantum task, it is necessary to determine to which “equivalent” entanglement groups the two multipartite entangled states belong. There are two frequently used equivalence relations that give rise to different classifications, i.e., local unitary (LU) equivalence and equivalence under local operations and classical communication (LOCC)\,\cite{PhysRevLett.104.020504m,PhysRevA.108.022220m,Commun328m}. For the former, to get a feeling for the classification from LU, it is instructive to compare it to the bipartite case, from which one concludes that two bipartite pure states are LU-equivalent if and only if their Schmidt coefficients coincide. This is a very satisfactory property for several reasons. First, it gives a concise answer to the entanglement classification problem. Second, the Schmidt coefficients that are exactly the set of eigenvalues of each of the reduced density matrices, have a simple physical meaning and can be estimated physically, e.g. using quantum state tomography or direct spectrum estimation methods\,\cite{PhysRevA.64.052311m,Commun261m}. Third, there is a simple and instructive proof of the validity of the Schmidt form involving just linear algebra. However, none of these three desirable properties can be generalized to systems with more than two subsystems. In this scenario, the multipartite pure states are mathematically described by tensors, and the well-defined matrix norms in bipartite case cannot be applied, makeing the physics behind it ambiguous. For the latter, as a more general equivalence relation, LOCC is usually described in the ``distant laboratories model", and allows classical propagation of measured results among $N$ laboratories where we have assumed that each particle has its own laboratory.  Such a protocol do not increase extra entanglement and hence also gives a justified proposal in ME classification. As a consequence, it concludes that two pure states $\ket{\psi}$ and $\ket{\phi}$ are equivalent under stochastic LOCC (SLOCC) if an invertible local operator relating them exists\,\cite{PhysRevA.63.012307m}. This result immediately indicates that the entanglement of any two qubits pure state is asymptotically equivalent, under deterministic LOCC, to that of the Einstein-Podolsky-Rosen state $(1/\sqrt{2})(\ket{00}+\ket{11})$. However, this simple entanglement classification is no longer true for $N > 2$-partite quantum states\,\cite{PhysRevA.62.062314m,Commun261m}. Surprisingly, one can obtain six inequivalent classes of states under SLOCC , even for the simplest case of three qubits pure states as shown in Fig.\,\ref{figS4}(background-color: green). More importantly, at the top of the hierarchy we can find two inequivalent classes of GME, i.e., Greenberger-Horne-Zeilinger (GHZ) state and W state. That is, for instance, it is impossible to convert a state belonging to GHZ class into W class under deterministic LOCC. Accordingly, multipartite entangled states, that are grouped into the same hierarchy, are suitable for implementing the same tasks of quantum information theory. Furthermore, the classification of mixed multipartite states becomes more complicated than the one for pure states and several more sophisticated tools are required. In this regime, great efforts have been made to show that the entanglement classification of mixed multipartite systems differs genuinely from the bipartite case. For example, a representative classification protocol, utilizing specifying compact convex subsets of the space of all states\,\cite{PhysRevLett.87.040401m,PhysRevA.67.012108m}, gives rise to four general classes of tripartite states, as shown in Fig.\,\ref{figS4} (background-color: green). In particular, witness operators are given to distinguish states from different classes. And again, the grouped classes are invariant under LU or SLOCC.

Secondly, in terms of topological entanglement, ME possesses more intricate geometric structures compared to BE. As mentioned earlier, three qubits have two in-equivalent SLOCC classes. The complexity, however, will increase greatly for $N>3$ qubits with the infinitely many classes\,\cite{PhysRevA.62.062314m, PhysRevA.65.052112m}.  Fortunately, a great deal of efforts are made to alleviate this problem, among which a novel perspective, i.e., dissipation-resistant rings formalism (DRRF), successfully links the entanglement classification of multipartite entangled states with their intrinsic topological properties \,\cite{Kauffman_2002m, PhysRevA.98.062335m, PhysRevA.98.062335m, PhysRevD.107.126005m}. This entirely new viewpoint was first proposed by \,\cite{aravind1997quantumm} and is further advanced in\,\cite{Kauffman2019m}.  It treats $N$-partite systems as $N$ closed rings, and the existence of ME indicates that the rings are connected. DRRF aims to analyze the robustness of multipartite entangled states to particle loss from which several $m$-resistant states are identified. A quantum state is $m$- resistant if the entanglement of the reduced state of $N-m$ subsystems is fragile with respect to the loss of any additional subsystem\,\cite{PhysRevA.100.062329m}. To understand the geometric classification from DRRF, it is instructive to compare it to the bipartite case.  In this sense, two rings can be connected in only one way, although in principle it can be done in infinitely many ways from the knot theory. The connection patterns would be richer and more interesting in multipartite quantum systems. In this context, the geometric classification of $N$-multipartite mixed
states theoretically starts from polynomial algebra and is more intricate for pure ones. For the former, one can find, taking $N=3$ as an example, two in-equivalent tripartite mixed state, i.e., 0-resistant and 1-resistant states. Their entanglement properties under partial trace could be represented vividly in two types of topological links as depicted in Fig.\,\ref{figS4} (background-color: mauve). In this diagrammatic representation, one can suddenly capture the following interesting fact: After removing any of the rings (a trace operation in physics), the connections of the three rings are broken or remain connected corresponding to 0-resistant (Borromean link) and 1-resistant (Hopf link) states, respectively.  For the latter, one can identify all of the m-resistant pure qubit states from different approaches, among which stellar representation of Majorana \,\cite{Aulbach_2010m, PhysRevA.81.062347m, PhysRevA.85.032314m} allows one to ascribe to entanglement a degree of geometrical intuition. Any $N$-qubit pure state can be well defined using $N$ pairs of Majorana points $(\theta_{j},\varphi_{j})$ in the Bloch sphere where each point looks like a star in the sky\,\cite{PhysRevA.81.062347m}. In this way, we are able to convert the problem of topological entanglement classification into counting the number of stars located at the North Pole. As shown in Fig.\,\ref{figS4} (background-color: mauve), one can classify two different types of tripartite pure states from the constellations representation resulting in 0-resistant (0 star in North Pole) and 1-resistant (1 star in North Pole) states, that are exactly correspond to the GHZ and W states. To a certain extent, the distance between the stars paves a new route for quantification of ME as the fully separate states correspond to all Majorana points pointing towards the North Pole, although such a criterion is not uniquely defined. Again, we argue that such a geometric interpretation for multipartite entangled states is genuinely distinct from that of bipartite case, even for the simplest scenario consisting of three subsystems.

Thirdly, in terms of entanglement measurement, the quantification of ME proves to be quite challenging to compute and conceptually difficult to explain in comparison with the BE that has been well understood and quantified\,\cite{RevModPhys.81.865m}.  We may thus need to come to terms with the fact that a canonical theory of ME may not exist. That is, as we will recall below, despite a basically unique way of quantifying bipartite state entanglement, for example, by concurrence\,\cite{PhysRevLett.78.5022m}or logarithmic negativity\,\cite{PhysRevA.65.032314m}, the “right” multi-partite state entanglement measure strongly depends on the intended use of the entangled states. More specifically, there is no unique way to quantify entanglement for three or more subsystems, as different entanglement measures induce different orderings and are maximized by different states\,\cite{PhysRevA.100.062329m}. As an example, the three-qubit GHZ state is the most entangled with respect to an important three-party entanglement measure called the three-tangle (or residual tangle), while the W state maximizes the average entanglement contained in two-party reductions. Therefore, it is reasonable to characterize ME depending on the applications considered. Furthermore, it has been demonstrated that quantifying entanglement in mixed multipartite states is difficult to generalize straightforwardly from that in pure multipartite entangled states, which remains a significant challenge today. For example, the quantification of mixed multipartite entangled states from the strategy of three-tangle is fundamentally undefined and extremely impractical on computational aspects since it is required to find all of the possible pure state decompositions for a mixed multipartite density matrix\,\cite{PhysRevA.61.052306m}. Unfortunately, the decomposition of a given multipartite density matrix is not unique! Despite these difficulties, several criteria have been proposed that are highly efficient in detecting ME or even GME, such as correlation tensors \,\cite{PhysRevA.84.062306m, Eur.Phys.J.Plusm, PhysRevA.96.052314m}, concurrence triangle \,\cite{PhysRevLett.127.040403m, JIN2023106155m, PhysRevLett.132.151602m}, quantum Fisher information   (QFI)\,\cite{PhysRevA.94.020101m, PhysRevApplied.18.024065m} and higher order entanglement witnesses for non-Gaussian multipartite entangled states\,\cite{PhysRevLett.130.093602m}. In particular, each criterion carries its unique benefits and intended preference in detecting ME. For instance, the criterion based on the correlation tensors maintains the well-established characteristics of standard norms by applying a matricizations procedure. Moreover, QFI's criterion is intended to pinpoint metrologically beneficial ME.  This method of detecting ME offers significant advantages in entanglement classification as the bounds of QFI can be further beaten by increasing the number of entangled particles. However, we still have a long way to go in fully understanding ME, and even faithfully measuring and characterizing ME remain long-standing challenges in the context of quantum information theory.

\begin{figure*}
	\includegraphics[width=16cm]{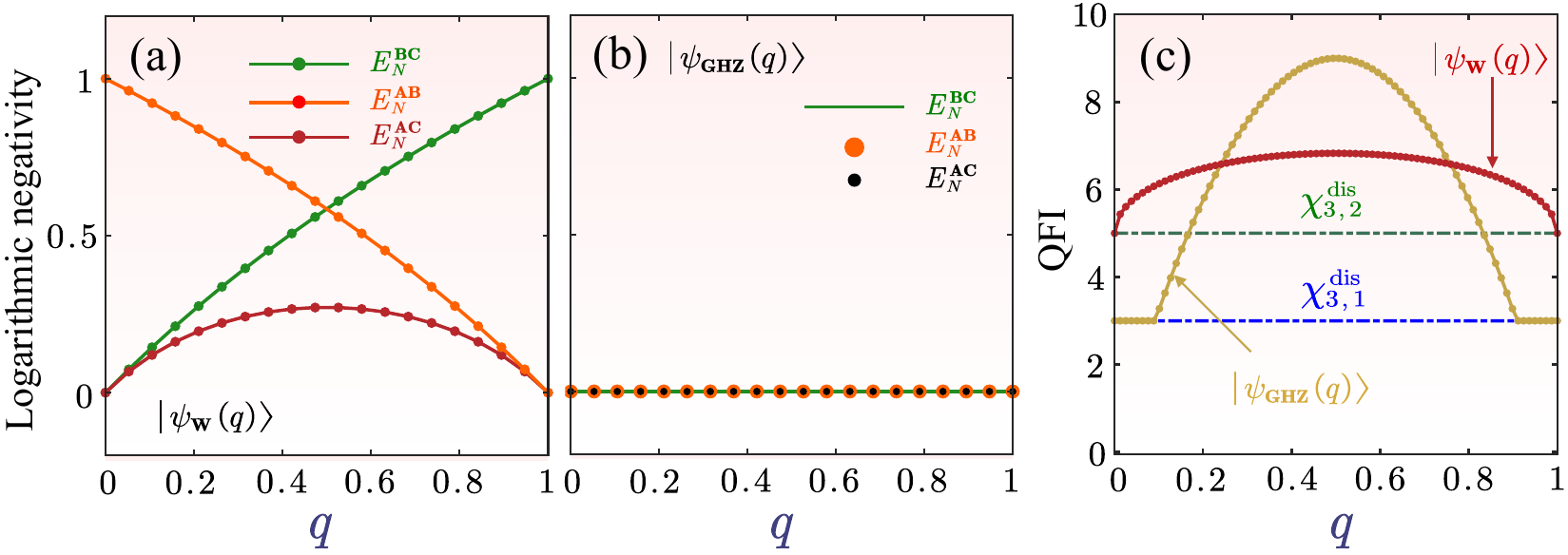}
	\caption{Logarithmic negativity for $q$-dependent reduced states (a) $\ket{\psi_{{\rm W}}}(q)$ and (b) $\ket{\psi_{{\rm GHZ}}}(q)$ that are plotted as a function of dimensionless parameter $q$. The corresponding measurements from QFI are plotted in (c), where $\chi^{{\rm dis}}_{3,1}$ and  $\chi^{{\rm dis}}_{3,2}$ are the bounds beyond which the state is entangled or genuinely entangled.}
	\label{figS5}
\end{figure*}

To further illustrate the non-trivial nature of measurements of ME, here we demonstrate that the delicate structure of the generated ME in our work can not be detected faithfully by the entanglement measures from BE. We note that the generated GME in our work may be equivalent to the GHZ class and the generated entangled multipartite cat states in the main text are typical GHZ states. Specifically, if we use bipartite entanglement criteria such as logarithmic negativity to assess a GHZ-class tripartite entangled state, we would conclude that it is an totally unentangled quantum state. The more detailed explanations are given as follows. On the one hand, we recall that the pure states belonging to GHZ category are 0-resistant. That is to say, the entanglement of the three subsystems (i.e., the connections of three rings) will be broken after tracing out any of the subsystems (i.e., removing any of the rings). When we use a bipartite entanglement criterion, such as logarithmic negativity or concurrence, to detect tripartite entanglement, an elimination operation is inevitably introduced. Since the generated GME in our work falls squarely into this category, we conclude that the simple usage of bipartite entanglement criteria fails to detect the GME in our work. To elaborate this fact in more detail, we present numerical simulations of ME detection by considering two following types of tripartite multipartite entangled states
\begin{align}\label{S56}
		\ket{\psi_{{\rm GHZ}}}(q)&=\sqrt{q}\ket{000}+\sqrt{1-q}\ket{111},\nonumber\\
\ket{\psi_{{\rm W}}}(q)&=\sqrt{q}\ket{001}+\sqrt{1-q}\ket{100}+\ket{010}.
\end{align}
As shown in Fig.\,\ref{figS5}(a), we plot the logarithmic negativity of reduced bipartite states by tracing out $A$, $B$ and $C$ in tripartite state $\ket{\psi_{{\rm W}}}(q)$, with the measured results are labeled by $E^{{\rm BC}}_{N}$ (green dotted line), $E^{{\rm AC}}_{N}$ (brown dotted line) and $E^{{\rm AB}}_{N}$ (orange dotted line), respectively. As anticipated, the logarithmic negativity can, to some extent,  detect maximal BE for partitions $AB$ and $BC$ when $q$ equals to $0$ and $1$, respectively. This also recalls the fact that the pure states belonging to W category are 1-resistant and maximize the average entanglement contained in two-party reductions. However, all the entanglement correlations between these partitions are broken in the case of $\ket{\psi_{{\rm GHZ}}}(q)$ [see Fig.\,\ref{figS5}(b)]. In this situation, we detect nothing but 0 from the bipartite entanglement criterion! In order to measure the existing ME more faithfully, we then choose QFI as our ME witness as shown in Fig.\,\ref{figS5}(c). We find the QFI can perfectly detect and also classify states $\ket{\psi_{{\rm W}}}(q)$  as genuine tripartite entanglement, except for $q=0$ and $q=1$ that correspond to bipartite entanglement. More importantly, it also successfully detects the ME for states $\ket{\psi_{{\rm GHZ}}}(q)$ in a wide range of parameters. Note that the states $\ket{\psi_{{\rm GHZ}}}(q)$ are unentangled for $q=0$ and $q=1$. Also, the measuring results between $\chi_{3,1}$ and  $\chi_{3,2}$ do not imply that the states $\ket{\psi_{{\rm GHZ}}}(q)$ are bipartite, but rather that the states contain at least bipartite entanglement. Interestingly, the QFI seems insensitive to the states within the regions $0\le q\le (1-\sqrt{(N-1)/N})/2$ or $1\ge q\ge (1+\sqrt{(N-1)/N})/2$, recalling that it only picks out  metrologically useful ME\,\cite{PhysRevA.82.012337m}.

On the other, it may seem like common sense that tripartite states are fully inseparable if non-zero logarithmic negativities are detected for all three partitions $AB$, $AC$, and $BC$. However, this is an illusion and thus it is impossible to detect ME faithfully only from bipartite entanglement criteria. For example, if a given state could be characterized as a mixture
\begin{align}\label{S57}
 \hat{\rho}=P_{1}\sum\limits_{i}\mathcal{F}_{i}^{(1)}\hat{\rho}^{i}_{1,23}+P_{2}\sum\limits_{j}\mathcal{F}_{j}^{(2)}\hat{\rho}^{j}_{2,13}+P_{3}\sum\limits_{k}\mathcal{F}_{k}^{(3)}\hat{\rho}^{k}_{3,12},
\end{align}
where $\sum\limits^{3}_{i=1}P_{i}=1$ , $\sum\limits^{3}_{i=1}\mathcal{F}_{i}^{(j)}=1$ , and $\hat{\rho}_{i,jk}$ is denoted as $\sum_{\gamma}\eta_{\gamma}\hat{\rho}^{\gamma}_{i}\otimes\hat{\rho}^{\gamma}_{jk}$ for brevity with $\sum_{\gamma}\eta_{\gamma}=1$, we can obtain $E^{XY}_{N}\neq 0$  for any $X,Y\in \{A,B,C\}$ and $X \neq Y$ when $P_{i}\neq 0$. However, the entanglement structure of  form Eq.\,(\ref{S57}) is truly biseparable and conceptually not the genuine ME.

Last but not least, in terms of entanglement application, the ME offers significant advantages in various quantum tasks beyond the BE, which are closely associated with the subtle properties mentioned above. ME serves as a remarkable resource in quantum information science, acting as a fundamental component in one-way quantum computing \,\cite{PhysRevLett.86.5188m, PhysRevLett.99.120503m, PhysRevLett.101.130501m}, multipartite teleportation and dense coding\,\cite{PhysRevLett.96.060502m}, and GME-based quantum secrete sharing\,\cite{natphys167m}. Moreover, in the domain of multipartite continuous variable entanglement, non-Gaussian states and gates play essential or advantageous roles in various applications, including quantum-enhanced sensing\,\cite{science1250147m}, entanglement distillation\,\cite{PhysRevLett.89.137903m, science0070m}, and in particular to universal quantum computation\,\cite{PhysRevLett.97.110501m, PhysRevLett.109.230503m}. Importantly, in our current letter, we have successfully generated non-Gaussian genuine multipartite entangled states in a deterministic manner. These states show a high level of resilience to the surrounding environment.
%\bibliography{referSM}
\end{document}